\documentclass{ptephy_v1}%%%%where ptephy_v1 is the template name

\preprintnumber{XXXX-XXXX} %%% %%% Insert preprint number here
\usepackage{hyperref}
\usepackage{aas_macros} % for mentioning MNRAS
\usepackage{amsmath} %for dealing with mathematics,
\usepackage{amsthm} %for dealing with theorem environments,
\usepackage{hyperref} %for linking the cross references
\usepackage{graphics} %for dealing with figures.
\usepackage{algorithmic} %for describing algorithms
\usepackage{subfig} %for getting the subfigures e.g., "Figure 1a and 1b" etc.
\usepackage{url} %It provides better support for handling and breaking URLs.
\usepackage{natbib}
\usepackage{hyperref}

\usepackage{bm}

\begin{document}

\title{Systematic local simulations of fast neutrino flavor conversions with scattering effects}

%%%% To generate auto affiliation numbers please use \author{}\affil{} command

\author[1,2,3,4]{Milad Delfan Azari}
\affil[1]{Department of Physics, Faculty of Science and Engineering, Waseda University, 3-4-1 Okubo, Shinjuku, Tokyo 169-8555, Japan} 
\affil[2]{Research Institute for Science and Engineering, Waseda University, 3-4-1 Okubo, Shinjuku, Tokyo 169-8555, Japan}
\affil[3]{Department of Mathematics, Temple University, Japan Campus, 1-14-29 Taishido, Setagaya, Tokyo 154-0004, Japan}
\affil[4]{Department of Physics, Temple University, Japan Campus, 1-14-29 Taishido, Setagaya, Tokyo 154-0004, Japan
\email{milad@heap.phys.waseda.ac.jp}}

\author[5]{Hirokazu Sasaki}
\affil[5]{Los Alamos National Laboratory, Los Alamos, New Mexico 87545, USA}

\author[6]{Tomoya Takiwaki}
%\author[3]{} %%% Use optional bracket [3] to change the respective address
\affil[6]{National Astronomical Observatory of Japan, 2-21-1 Osawa, Mitaka, Tokyo 181-8588, Japan}

\author[7,8]{Hirotada Okawa}
%\thanks{These authors contributed equally to this work}}
\affil[7]{Waseda Institute for Advanced Study, Waseda University, Tokyo 169-0051, Japan}
\affil[8]{Faculty of Software and Information Technology, Aomori University, Tokyo 2-10-1, 134-0087, Japan}

%%% To include the collaborator name... Please use the command "\collaborator"
%%% For example: \collaborator{ATLAS Collaboration}

\begin{abstract}%
We investigate the dynamics of fast neutrino flavor conversions (FFCs) in the one-dimensional (1D) and zero-dimensional (0D) models, in which spatial advection is considered and ignored, respectively. In this study, we employ snapshots obtained by our self-consistent, realistic Boltzmann-neutrino-radiation-hydrodynamics simulations. We show that the FFC growth rate is considerably larger in the 1D model than in the 0D model, as expected from the previous linear analysis results. We find that the momentum space dimension does not significantly influence the neutrino transition probability in 1D models. On the other hand, in the 0D model without collisions, the FFC depends on the momentum space, and the azimuthal angle dependence breaks the periodicity of the FFC. Our study demonstrates that collisional instability can lead to further flavor conversions on a long time scale in 1D models after the asymptotic state of FFC has been reached. Such an effect should be taken into consideration when the fast and collisional flavor instabilities coexist.
\end{abstract}

\subjectindex{E26, E45}

\maketitle

\section{Introduction}

Neutrinos are the most abundant massive particles in the universe \cite{PhysRevD.52.3184}. Although they are massive, their masses are much smaller than those of their charged lepton counterparts. Due to the mismatch between their flavor and mass eigenstates, they can experience flavor transformation when propagating through vacuum \cite{Super-Kamiokande:1998kpq, PhysRevD.78.083007}. Weak interactions with surrounding matter can change the dispersion relation in vacuum and induce resonant conversions known as the MSW (Mikheyev-Smirnov-Wolfenstein) effect \cite{PhysRevD.17.2369}. In a dense neutrino environment, due to the high population of neutrinos, interaction with other neutrinos generates their own self-energy, which can lead to collective neutrino oscillations \cite{Duan2010,Mirizzi2016,Tamborra2021,Capozzi2022review,Richers2022review,Volpe2023review}.

Massive stars ($\textrm{M}_{\star}\gtrsim \textrm{8\,M}_{\odot}$) at the end of their lives may explode as core-collapse supernovae (CCSNe). During this event, about $10^{53}$ erg energy is released in an enormous amount of neutrinos in all flavors ($\nu_e$, $\nu_{\mu}$ and $\nu_{\tau}$). It is believed that these neutrinos are essential ingredients in CCSNe's explosion mechanism. If a fraction of $\nu_e$ and $\bar\nu_e$ are reabsorbed by the matter between the shock front and the so-called gain radius and deposit their energy to push the stagnated shock again, CCSNe are expected. It is known that $\nu_{\mu}$, $\nu_{\tau}$ and their anti-particles have higher average energies than $\nu_{e}$ and $\bar\nu_{e}$ due to the weak interactions of $\nu_{\mu}$, $\bar\nu_{\mu}$, $\nu_{\tau}$  and $\bar\nu_{\tau}$ with matters while propagating. If the former neutrinos ($\nu_{\mu}$, $\bar\nu_{\mu}$, $\nu_{\tau}$  and $\bar\nu_{\tau}$) are converted to the latter neutrinos ($\nu_{e}$ and $\bar\nu_{e}$), those converted neutrinos will be absorbed by matter. Such neutrino flavor conversion transfers more neutrino energy to the matter behind the shock, resulting in either a successful explosion or a failure of the shock revive \cite{Suwa_2011,Ehring2023a,Ehring2023b,Nagakura2023may}.
Therefore, collective neutrino oscillations have attracted the most attention since about a decade ago. However, their mechanisms are still unknown due to their non-linearity and complex nature.

In particular, among collective neutrino oscillations, the fast flavor conversion (FFC) has been speculated to occur deep inside CCSNe. The frequency is proportional to the neutrino potential $\mu \sim\sqrt 2 G_{F}n_{\nu}$ \cite{PhysRevD.79.105003, PhysRevLett.116.081101, PhysRevD.72.045003} and high enough to ignore the vacuum frequency. The linear stability analysis was used to reveal the growth of the FFC before solving the non-linear evolution. It has been pointed out that FFC occurs as a result of the crossing between the angular distribution of electron-type neutrinos ($\nu_e$) and anti-electron neutrinos ($\bar\nu_e$), the so-called electron-lepton number (ELN) crossing \cite{Johns2021jun,Dasgupta2022}. The possibility of FFC in CCSNe has been studied intensively as post processes using sophisticated Boltzmann neutrino radiation hydrodynamic simulations \cite{PhysRevD.99.103011, PhysRevD.101.023018,Abbar2019,Abbar2020feb,Abbar2020may,Abbar2021}. More simplified two moment scheme can be used to find ELN crossing with reconstruction technique of neutrino angular distribution \cite{Johns2021jun,Nagakura2021jun,Nagakura2021sep,Abbar2023may,Abbar2024}.
Even though the fact that angular crossings in neutrino-flavor-lepton-numbers (NFLN), which are necessary and sufficient conditions for FFC, are unlikely to occur in spherically symmetric CCSN models, they have been observed more frequently in multidimensional models~\cite{Shalgar2019,Nagakura2021}. 

Many simulations have recently been performed to understand the non-linear evolution beyond the linear regime, solving the time evolution of FFC.
Most of them treat small-scale local areas \cite[e.g,][]{Dasgupta2018,Capozzi2020,Johns2020,Kato2020,Kato2021,Meng-Ru2021,Richers2021,Shalgar2022}.
Several large-scale global simulations are performed with the parameter re-scaling technique \cite{Nagakura2022dec,Nagakura2023mar,Xiong2023}.
The effect of collisions has also been investigated in terms of their effect on FFC \citep{Cirigliano2017,Capozzi2019}. Some papers reported that neutral-current collisions enhance fast flavor conversions (e.g., \cite{Shalgar2021,Sasaki2022,Johns2022}), while others have reported damping effects \cite{Martin2021,Sigl2022,Padilla-Gay2022}. Although these previous studies focused on the energy-integrated neutrino angular distribution, energy-dependent scattering, emission, and absorption are also studied \cite{Kato2022,Kato2023,Lin2023}. Ref.~\cite{Johns2022} pointed out the inconsistency of the initial profile and the setting of the numerical simulation.  A similar but different mechanism, collisional instability, is also found and discussed so far  \cite{Johns2023,Kato2023,Liu2023}. Some of the results seem controversial with each other, and the overall role of collisions in fast flavor conversion is still unclear.

The goal of this paper is to investigate FFCs in a non-linear regime and the effect of neutrino-nucleon scatterings by examining the dynamics of four flavors of neutrinos ($\nu_e, \bar\nu_e$, $\nu_x$, and $\bar{\nu}_x$) in both zero-dimensional (0D) and one-dimensional (1D) setups. In  zero-dimensional (0D) models, we ignore the advection term in the transport equations.
The dimension of momentum spaces is systematically changed, and the initial condition is taken from the realistic two-dimensional (2D) fully self-consistent Boltzmann-neutrino-radiation-hydrodynamics simulations.

This article is organized as follows. In Section~\ref{sec:formulation}, we introduce the equations that are used in our analysis. Section~\ref{sec:models} presents detailed information on our models. In Section~\ref{sec:results}, we show our results, and finally, in Section~\ref{sec:summary}, we summarize our results and conclude the paper.

\section{Formulation}\label{sec:formulation}

Similar to the previous works \cite{PhysRevD.62.093026, PhysRevD.89.105004, PhysRevD.103.063002, PhysRevD.94.033009,Sasaki2022}, we calculate fast neutrino flavor conversions of $\nu_e$ and $\nu_x$ with their anti-particles $\bar\nu_e$ and $\bar\nu_x$ considering the collision effects of neutrino scattering. We use {\small QDSCNO} code \cite[Sasaki and Takiwaki (2022), hereafter we call this S22,][]{Sasaki2022} to study the neutrino flavor conversions based on the neutrino density matrices. The evolution of neutrinos is described as
\begin{align}
\partial_t \rho + v_i \partial_i\rho =-i \left[H,\rho\right]+C\left(\rho, \bar{\rho}\right),\label{eq:QKEneu}\\
\partial_t \bar{\rho} + v_i\partial_i\bar{\rho} =-i \left[\bar{H},\bar{\rho}\right]+\bar{C}\left(\rho, \bar{\rho}\right),\label{eq:QKEant}
\end{align}
where $\rho$, $H$ and $C$ are density matrix, Hamiltonian, and collision terms for neutrinos, respectively. 
$\bar{\rho}$, $\bar H$ and $\bar C$ correspond to those for anti-neutrinos. The velocity of the neutrino for $i$-direction is denoted by $v_i$. In this study, we consider one spatial dimension $z$ and $v_z=c\cos\theta_\nu$, where $c$ is the speed of light. Here the neutrino distribution is assumed homogeneous in the $x$ and $y$ direction.

In this article, for simplicity, we use two flavor approximation. The density matrix consists of independent components, $\rho_{ij},\bar{\rho}_{ij}$, where $i,j=e, x$.
Note that those are Hermitian matrix, i.e., $\rho_{ij} =\rho_{ji}^*$, where $*$ is complex conjugate.
The initial density matrix is normalized by the neutrino distribution function, and can be written as 
\begin{align}
\rho_{ii}(E_{\nu},\theta_{\nu}, \phi_{\nu}) =& 4\pi  f_{\nu_i}(E_{\nu},\theta_{\nu}, \phi_{\nu})  /  n_{E,\nu_{e}},\\
n_{E,\nu_{e}} (E_{\nu})=& \int {\rm d}\Omega_\nu f_{\nu_{e}}, % 
\end{align}
where ${\rm d}\Omega_\nu=\sin\theta_{\nu}{\rm d}\theta_{\nu}{\rm d}\phi_{\nu}$. $E_{\nu}$ is the energy, and $\theta_{\nu}$ and $\phi_{\nu}$ are angles in the momentum space and $f_{\nu_i}$ denotes the distribution function of $\nu_i$.
Here we consider a three-dimensional momentum space. 
In some models, we reduce the number of dimensions and the definition for one- or two-dimensional momentum space is explained in Appendix~\ref{sec:normalization}.

The left hand sides in Eqs.~\eqref{eq:QKEneu} and \eqref{eq:QKEant} are solved by 5th-order WENO method. 
For the time-integration, Runge-Kutta 4th method is used.
This part is not taken into account in the previous version of {\small QDSCNO}~\cite{Sasaki2022}.
For the details of WENO, see 
{\small COSE$\nu$}~\cite{George2023cosen} and
{\small GRQKNT}~\cite{Nagakura2022GRQKNT}.
A code comparison is performed in Ref.~\cite{Richers2022}.
We also test the problem in Ref.~\cite{Richers2022}, which is presented in Appendix~\ref{sec:codeverification}.

The Hamiltonian in Eqs.~\eqref{eq:QKEneu} and \eqref{eq:QKEant} is written as
\begin{equation}
H = H_{\text{vacuum}}+H_{\text{matter}}+H_{\nu\nu},
\end{equation}
where the vacuum term is given as 
\begin{equation}
H_{\text{vacuum}} = \frac{\Delta m^2}{4E_\nu}
\left(
\begin{array}{cc}
 -\cos\theta_{\rm v}    &   \sin\theta_{\rm v} \\
   \sin\theta_{\rm v}    &   \cos\theta_{\rm v} \\
\end{array}\right),\label{eq:hvac}
\end{equation}
where $\Delta m^2 =2.5 \times 10^{-6}\,{\rm eV}^2$.
Note that we impose an effective vacuum mixing angle $\theta_{\rm v}=10^{-6}$ instead of the matter potential.

The collective part is given as 
%\newpage
\begin{equation}
H_{\nu\nu}(\theta_{\nu},\phi_\nu)=
\int {\rm d}{\mu_E^\prime}
\int \frac{{\rm d}\phi_\nu^\prime}{2\pi}\int^{1}_{-1}\frac{\mathrm{d}\cos\theta_\nu^{\prime}}{2}h_{\nu\nu},
\end{equation}
where ${\rm d}\mu^\prime_{E}$ is
\begin{equation}
{\rm d}\mu^\prime_E=\sqrt{2}G_F \frac{E^{\prime^2}_\nu{\rm d}E_\nu^\prime}{\left(2\pi\hbar c\right)^3}n_{E,\nu_{e}}, 
\end{equation}
and $h_{\nu\nu}$ is 
\begin{equation}
\begin{split}
h_{\nu\nu} =&[\rho(E^\prime_\nu,\theta_{\nu}^\prime,\phi_{\nu}^\prime)-\bar{\rho}(E^\prime_\nu,\theta_{\nu}^\prime,\phi_{\nu}^\prime)] \\
&\times \left[1-\cos\theta_{\nu}\cos\theta_{\nu}^\prime -\sin\theta_{\nu}\sin\theta_{\nu}^\prime\right.\\
&
\times\left.
(\cos\phi_{\nu}\cos\phi_{\nu}^\prime+\sin\phi_{\nu}\sin\phi_{\nu}^\prime)\right].
\end{split}\label{eq:hamiltonian}
\end{equation}

The collision terms in Eqs.~\eqref{eq:QKEneu} and \eqref{eq:QKEant} are similar to the Eqs.~(2) and (8) in Ref.~\cite{Johns2023} (see also Refs.~\cite{Martin2021,Sasaki2022}).
Considering the collision terms of neutrino scattering in flavor-blind Neutral-Current (NC) reactions and employing the elastic neutrino-nucleon collisions, we have
\begin{align}
C^{\mathrm {NC}}(\rho) =  -\kappa_{0}\rho
+\int\frac{{\rm d}\Omega_\nu^\prime}{4\pi}\left(\kappa_{0}-\frac{\kappa_{1}}{3}\cos\theta_\nu\cos\theta^{\prime}_\nu\right)\rho^{\prime},\label{eq:NCreacrn}\\
\bar C^{\mathrm {NC}}(\bar{\rho}) =  -\kappa_{0}\bar{\rho}
+\int\frac{{\rm d}\Omega_\nu^\prime}{4\pi}
\left(\kappa_{0}-\frac{\kappa_{1}}{3}\cos\theta_\nu\cos\theta^{\prime}_\nu\right)\bar{\rho}^{\prime},\label{eq:NCreacra}
\end{align}
where $\rho^{\prime}=\rho(E_\nu,\theta_{\nu}^\prime,\phi_{\nu}^\prime)$ and $\bar\rho^{\prime}=\bar{\rho}(E_\nu,\theta_{\nu}^\prime,\phi_{\nu}^\prime).$ Here, the non-axisymmetric scattering is ignored. The coefficients in the NC-collision terms are given by \cite{Martin2021}
\begin{align}
\kappa_{0} = \frac{3G^2 E_\nu^2}{\pi}\sum_{N=n,p}n_{N}\left\{
(c_{A}^{N})^{2}+\frac{(c_{V}^{N})^{2}}{3}
\right\},\label{eq:kappa0}\\
\kappa_{1} = \frac{3G^2 E_\nu^2}{\pi}\sum_{N=n,p}n_{N}\left\{
(c_{A}^{N})^{2}-(c_{V}^{N})^{2}
\right\},\label{eq:kappa1}
\end{align}
where $n_{N}(=n_{n},n_{p})$ is the nucleon number density inside the matter.

We also consider the effect of flavor-resolving Charged-Current (CC) reactions by employing the collision term of neutrino-electron scatterings ~\cite{Johns2023},
\begin{align}
C^{\mathrm{CC}}(\rho) = -\Gamma^{\mathrm{CC}}
\begin{pmatrix}
\rho_{ee} & \frac{\rho_{ex}}{2} \\
\frac{\rho_{xe}}{2} & 0
\end{pmatrix}
+\Gamma^{\mathrm{CC}}
\begin{pmatrix}
%\langle \rho_{ee} \rangle & 0 \\
\int\frac{{\rm d}\Omega_\nu^{\prime}}{4\pi}\rho_{ee}^{\prime}& 0 \\
0 & 0
\end{pmatrix},\label{eq:CCreacrn}\\
\bar C^{\mathrm{CC}}(\bar\rho) = -\bar\Gamma^{\mathrm{CC}}
\begin{pmatrix}
\bar\rho_{ee} & \frac{\bar\rho_{ex}}{2} \\
\frac{\bar\rho_{xe}}{2} & 0
\end{pmatrix}
+\bar\Gamma^{\mathrm{CC}}
\begin{pmatrix}
%\langle \bar\rho_{ee} \rangle & 0 \\
\int\frac{{\rm d}\Omega_\nu^{\prime}}{4\pi}\bar\rho_{ee}^{\prime}& 0 \\
0 & 0
\end{pmatrix},\label{eq:CCreacra}
\end{align}
where the $\Gamma^{\mathrm {CC}} = 1/\lambda_{\nu_e e}$,  $\bar\Gamma^{\mathrm {CC}} = 1/\lambda_{\bar\nu_e e}$ and mean free path $\lambda$ are calculated from the net electron density $n_e$, and the neutrino scattering cross section $\sigma_{\nu_\alpha e}$ \cite{Johns2023}.
\begin{eqnarray}
\Gamma^{\mathrm {CC}}=\frac{1}{\lambda_{\nu_e e}}&\sim& n_e(\sigma_{\nu_e e}-\sigma_{\nu_x e}), \label{eq:ESCCn}\\
\bar\Gamma^{\mathrm {CC}}=\frac{1}{\lambda_{\bar\nu_e e}}&\sim& n_e(\sigma_{\bar\nu_e e}-\sigma_{\bar\nu_x e}). \label{eq:ESCCa}
\end{eqnarray}
In Eq.~\eqref{eq:ESCCn}, we extract the CC part, subtracting the NC part, $n_e \sigma_{\nu_x e}$, from the total scattering rates, $n_e \sigma_{\nu_e e}$, and same for anti-neutrinos in Eq.~\eqref{eq:ESCCa}.
In the above equation, 
\begin{eqnarray}
\sigma_i = \frac{3}{8}\sigma_0 c_{i}\left(k_{\rm B}T+\frac{\mu_e}{4}\right)\frac{E_\nu}{(m_e c^2)^2},
\end{eqnarray}
where $c_{i} = 2.333$, $1.0$, $0.3$ and $0.3$ for $\nu_e$, $\bar{\nu}_e$, $\nu_x$ and $\bar{\nu}_x$, respectively~\cite{Bowers1982ApJS}.
The cross section depends on the temperature $T$, and the chemical potential of electron $\mu_e$. Here $\sigma_0 = 1.7\times 10^{-44}\,{\rm cm^2}$.
On this collision, the elastic approximation is not applicable \cite[e.g,][]{Rampp2002}, but we assume it for simplicity.
Ref.~\cite{Richers:2019grc} is helpful for finding more detailed equations that include emission and absorption.

\section{Models}\label{sec:models}

In this section, we provide detailed information on our models. 
The initial conditions in this article are provided by Delfan Azari et al. (2019)~\cite{PhysRevD.99.103011} and  Delfan Azari et al. (2020)~\cite{PhysRevD.101.023018}, hereafter we call these references D19 and D20. The first paper (D19) did not find any ELN crossing but the second paper (D20) obtained the ELN crossing in the protoneutron star (PNS) region. These results are favorable to our study and are the main motivation to conduct this research.

The initial profile is taken from the results of the realistic two-dimensional (2D) fully self-consistent Boltzmann-neutrino-radiation-hydrodynamics simulations for the progenitor model of non-rotating 11.2\,$\textrm M_{\odot}$ \cite{RevModPhys.74.1015} which were performed on the Japanese K-supercomputer~\cite{Nagakura_2018}. In these simulations, three neutrino species, $\nu_e$, $\bar\nu_e$, and $\nu_x$ are considered, and their distributions are computed on spherical coordinates $(r, \theta)$ under spatial axisymmetry. We employed spherical coordinates in momentum space $(E_\nu, \theta_{\nu}, \phi_{\nu})$, in which the two angles are measured from the local radial direction. The computational domain covers $0 \leq r \leq 5000\,{\rm km}$, $0\leq \theta \leq \pi$, $0 \leq E_\nu \leq 300\,{\rm MeV}$, 0 $\leq$ $\theta_{\nu}$ $\leq$ $\pi$ and 0 $\leq$ $\phi_{\nu}$ $\leq$ 2$\pi$ with 384~(r), 128~($\theta$), 20~($E_\nu$), 10~($\theta_{\nu}$) and 6~($\phi_{\nu}$) mesh cells. The paper adopts the Furusawa-Shen equation of state (FSEOS) that is based on relativistic mean field theory for nuclear matter \cite{Furusawa_2013}.

In this article, we present only the results of our analysis at the post-bounce time of $\textrm t_{\textrm {pb}} = 190$ ms. The reason for choosing this time is due to the fact that the crossing was found for the first time at a point where the radius $r = 16.5\,{\rm km}$ and the spatial zenith $\theta = 2.1\,{\rm rad}$ at the post-bounce time of $\textrm t_{\textrm {pb}} = 190$\ ms in D20~\cite{PhysRevD.101.023018}.

The hydrodynamic properties of this point is summarized as follows. The density $\rho_{\rm B}$, is $2.4\times 10^{13}\,{\rm g/cm^3}$; electron fraction $Y_e$, is 0.13; temperature $k_{\rm B} T$, is 20.4\,{$\rm MeV$}; and the chemical potential of electron $\mu_e$, is 56.9\,{$\rm MeV$}. The values are summarized in Table~\ref{tab:phys}.

\begin{table}[htb]
\centering
\caption{Summary of variables and values.}\label{tab:phys}
\begin{tabular}{p{13em}cr}
Variable & Symbol & Value \\
\hline
Baryon density & $\rho_{\rm B}$ & $2.4\times 10^{13}\,{\rm g/cm^3}$ \\
Electron fraction &  $Y_e$ & 0.13\\
Temperature & $T$ & $20.4\,{\rm MeV}$\\
Chemical potential of electrons & $\mu_e$ & $56.9\,{\rm MeV}$\\
Averaged neutrino energy & $\langle E_\nu \rangle$ & $64.0\,{\rm MeV}$\\
Potential of neutrino & $\mu$ & $1.93\times 10^{13}\,{\rm s^{-1}}$\\
Peak of dispersion relation & $\Omega_{\rm peak}$ & $3.9\times 10^{10}\,{\rm s^{-1}}$\\
Typical growth rate of $P_{\rm tra}$ in 1D model & $\Omega_{\rm 1D}$ & $\sim 8\times10^{10}\,{\rm s^{-1}}$\\
Typical growth rate of $P_{\rm tra}$ in 0D model & $\Omega_{\rm 0D}$ & $\sim 8\times10^{8}\,{\rm s^{-1}}$\\
Typical growth rate of $P_{\rm tra}$ in the late period in S1M1-LCC & $\Omega_{\rm S1M1-LCC}$ & $\sim1.5\times10^{6}\,{\rm s^{-1}}$\\
NC-collision rate & $\kappa_0$ & $0.43\times 10^{8}\,{\rm s}^{-1}$\\
NC-collision rate & $\kappa_1$ & $0.17\times 10^{8}\,{\rm s}^{-1}$\\
CC-collision rate for $\nu_e$ & $\Gamma^{\rm CC}$ & $0.058\times 10^{8}\,{\rm s}^{-1}$\\
CC-collision rate for $\bar{\nu}_e$ & $\bar{\Gamma}^{\rm CC}$ & $0.018\times 10^{8}\,{\rm s}^{-1}$\\
Growth rate of collisional instability& $\Omega_{\rm CI}$ & $0.406\times10^{6}\,{\rm s^{-1}}$\\
\hline
\end{tabular}
\end{table}
\begin{table*}[h]
\caption{In this table, we summarize the properties of the zero-dimensional models.
 The model names are written in the form of S and M, in which M stands for the dimension of momentum space, and S corresponds to the dimension of space. S0 represents zero-dimensional models. The meaning of the model name is also written in the description column. The numerical resolution of the simulation is summarized in $N_{\theta_\nu},N_{\phi_{\nu}},N_{E_\nu},N_z$ columns as well. Results column summarizes the feature of transition probability. S0M1 are the fiducial models in the zero-dimensional models. We show the feature of models comparing their evolution of the probability to that of the fiducial model. Finally, previous works using similar setup are summarized in References column.  
 }\label{tb:models}
\begin{tabular}{ |p{1.7cm}|p{4.4cm}|p{0.5cm}|p{0.5cm}|p{0.5cm}|p{0.3cm}|p{2.7cm}|p{1.2cm}|}
 \hline
\multicolumn{8}{|c|}{\small{Zero-dimensional models}} \\
\hline
 \small{Model} & \small{Description}& \small{$N_{\theta_{\nu}}$} & \small{$N_{\phi_{\nu}}$} & \small{$N_{E_\nu}$} & \small{$N_z$} &\small{Results}&\small{Ref.}\\
\hline
\small{S0M1}& \small{S0: 0D} & \small{200} & \small{1} & \small{1} & \small{1} & \small{Periodic, Fiducial} & \small{\cite{Shalgar2021,Sasaki2022}}  \\
\small{S0M1-NC}& \small{NC: Neutral-Current} & \small{200} & \small{1} & \small{1} & \small{1} &  \small{Suppressed} & \small{\cite{Shalgar2021}} \\
\small{S0M1-CC}& \small{CC: Charged-Current} & \small{200} & \small{1} & \small{1} & \small{1} &  \small{Enhanced} & \small{\cite{Sasaki2022,Johns2023}}\\
\small{S0M2E}& \small{M2E: Energy-dependent} & \small{100} & \small{1} & \small{30} & \small{1} &  \small{Not affected} & \small{\cite{Kato2022}}\\
\small{S0M2E-NC}&  & \small{100} & \small{1} & \small{30} & \small{1} &  \small{Enhanced} & \small{\cite{Kato2022}} \\
\small{S0M2E-CC}&  & \small{100} & \small{1} & \small{30} & \small{1} &   \small{Enhanced} & \\
\small{S0M2A}& \small{M2A: Azimuth-dependent} & \small{100} & \small{100} & \small{1} & \small{1} &  \small{Break periodicity} & \small{\cite{Shalgar2022}} \\
\small{S0M2A-NC}&  & \small{100} & \small{100} & \small{1} & \small{1} & \small{Suppressed} & \\
\small{S0M2A-CC}&  & \small{100} & \small{100} & \small{1} & \small{1} & \small{Enhanced} & \\
\small{S0M3}& \small{M3: 3D momentum space} & \small{100} & \small{100} & \small{30} & \small{1} &  \small{Break periodicity}& \\
\small{S0M3-NC}&  & \small{100} & \small{100} & \small{30} & \small{1} &  \small{Enhanced} & \\
\small{S0M3-CC}&  & \small{100} & \small{100} & \small{30} & \small{1} & \small{Enhanced} & \\
 \hline
\end{tabular}
\end{table*}
\begin{table*}[h]
\caption{In this table, we summarize the properties of the one-dimensional models.
 The model names are written in the form of S and M, in which M stands for the dimension of momentum space, and S corresponds to the dimension of space. S1 represents one-dimensional models. The meaning of the model name is also written in the description column. The numerical resolution of the simulation is summarized in $N_{\theta_\nu},N_{\phi_{\nu}},N_{E_\nu},N_z$ columns as well. Results column summarizes the feature of transition probability. S1M1 are the fiducial models in the one-dimensional models. We show the feature of models comparing their evolution of the probability to that of the fiducial model. Finally, previous works using similar setup are summarized in References column.  
 }\label{tb:models2_2}
\begin{tabular}{ |p{1.7cm}|p{4.7cm}|p{0.5cm}|p{0.4cm}|p{0.35cm}|p{0.55cm}|p{2.5cm}|p{1.1cm}|}
 \hline

\multicolumn{8}{|c|}{\small{One-dimensional models}} \\
\hline
 \small{Model} & \small{Description}& \small{$N_{\theta_{\nu}}$} & \small{$N_{\phi_{\nu}}$} & \small{$N_{E_\nu}$} & \small{$N_z$} &\small{Results}&\small{Ref.}\\
 \hline
\small{S1M1}& \small{S1: 1D} & \small{200} & \small{1} & \small{1} & \small{1024} & \small{Fiducial} & \small{\cite{Martin2021,Sigl2022}}\\
\small{S1M1-HP} & \small{HP: Homogeneous Perturbations} & \small{200} & \small{1} & \small{1} & \small{1024} &  \small{Suppressed} &  \\
\small{S1M1-NC} &  & \small{200} & \small{1} & \small{1} & \small{1024} & \small{Not affected} & \small{\cite{Martin2021,Sigl2022}}\\
\small{S1M1-CC} &  & \small{200} & \small{1} & \small{1} & \small{1024} & \small{Not affected} & \\
\small{S1M1-L}  & \small{L: Long-term}  & \small{200} & \small{1} & \small{1} & \small{1024} & \small{Fiducial} &  \\
\small{S1M1-LNC}&  & \small{200} & \small{1} & \small{1} & \small{1024} & \small{Slightly enhanced} & \\
\small{S1M1-LCC}&  & \small{200} & \small{1} & \small{1} & \small{1024} & \small{Enhanced} & \small{\cite{Johns2023}}\\
\small{S1M2E}   &  & \small{100} & \small{1} & \small{30} & \small{1024} & \small{Not affected} & \\
\small{S1M2E-NC}&  & \small{100} & \small{1}   & \small{30} & \small{1024} & \small{Not affected} & \\
\small{S1M2E-CC}&  & \small{100} & \small{1}   & \small{30} & \small{1024} & \small{Not affected} & \\
\small{S1M2A}   &  & \small{100} & \small{100} & \small{1}  & \small{1024} & \small{Not affected} & \\
\small{S1M2A-NC}&  & \small{100} & \small{100} & \small{1}  & \small{1024} & \small{Not affected} & \\
\small{S1M2A-CC}&  & \small{100} & \small{100} & \small{1}  & \small{1024} & \small{Not affected} & \\
\small{S1M3}    &  & \small{40}  & \small{40}  & \small{10} & \small{1024} & \small{Not affected} & \\
\small{S1M3-NC} &  & \small{40} & \small{40} & \small{10} & \small{1024} & \small{Not affected} & \\
\small{S1M3-CC} &  & \small{40} & \small{40} & \small{10} & \small{1024} & \small{Not affected}  & \\
 \hline
\end{tabular}
\end{table*}    
Using the initial profile, we systematically investigate the effects of dimension and collision. Tables \ref{tb:models} and~\ref{tb:models2_2} summarize our models. We perform zero- and one-dimensional simulations. The model names are written in the form of S and M. We call one-dimensional model S1 after its spatial dimension and zero-dimensional model S0 whose spatial dimension is zero.  The direction of $z$ is taken as $r$ direction in the background model.
For the momentum space, we have four choices, M1, M2E, M2A and M3 where their degree of freedom is $(\theta_\nu)$, $(E_\nu,\theta_\nu)$, $(\theta_\nu, \phi_\nu)$, and $(E_\nu, \theta_\nu, \phi_\nu)$, respectively.
In each model, remaining dimension is integrated (see Appendix~\ref{sec:normalization}).
In the model name, M stands for the dimension of momentum space. 

We also change the type of the scattering. In the model name, NC and CC are used for Neutral-Current and Charged-Current, see Eqs.~\eqref{eq:NCreacrn}--\eqref{eq:CCreacra}, for the description.

The initial noise level is controlled as follows.
\begin{align}
    \rho_{ex}=& (\rho_{ee}+\rho_{xx}) ( \epsilon_r+\epsilon_\imath \imath ),\\
    \bar{\rho}_{ex} =& ( \bar{\rho}_{ee}+ \bar{\rho}_{xx}) ( \bar{\epsilon}_r+\bar{\epsilon}_\imath \imath ),
\end{align}
where $\epsilon_r, \epsilon_\imath, \bar{\epsilon}_r,\bar{\epsilon}_\imath$ are the random numbers that are less than $10^{-8}$. Note that $\rho_{ex}$ and $\bar{\rho}_{ex} $ are complex numbers and $\epsilon_r, \epsilon_\imath$ are real numbers.
We impose this perturbation in all degrees of freedom. For example, in S1M1 model, different values of $\epsilon_r, \epsilon_\imath, \bar{\epsilon}_r,\bar{\epsilon}_\imath$ are used for each $z$ and $\theta_\nu$ grid.

\begin{figure} [h]
\includegraphics[width=.9\linewidth]{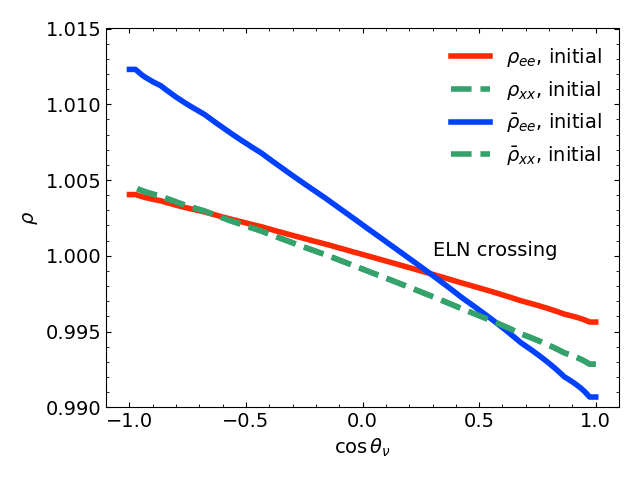}
\caption{Initial condition of M1 (energy-integrated axisymmetric) simulation. $\rho_{ee}$ (red), $\bar{\rho}_{ee}$ (blue), $\rho_{xx}$ and $\bar{\rho}_{xx}$ (green) correspond to the density matrix of $\nu_e$, $\bar\nu_e$, $\nu_x$ and $\bar\nu_x$, respectively.  Initially $\rho_{xx}=\bar{\rho}_{xx}$.
The position of electron-lepton number crossing is indicated by the text.
Appendix~\ref{sec:normalization} shows how to obtain M1 density matrix.
Note that our notation assumes $\int \frac{{\rm d}\cos\theta_\nu}{2} \rho_{ee} =1 $ and that would be factor 2 larger than some works that assume $\int {\rm d}\cos\theta_\nu \rho_{ee}  =1 $.}\label{fig:the-rho_ini}
\end{figure}

\section{Results}\label{sec:results}
We start showing our results by illustrating the initial condition of M1 (energy-integrated axisymmetric) profile in Figure~\ref{fig:the-rho_ini}. The density matrix of $\nu_e$, $\bar{\nu_e}$, $\nu_x$ is shown as red, blue, and green, respectively. Note that $\bar\nu_x = \nu_x$.
The position of electron-lepton number (ELN) crossing is indicated by the text and it occurs at $\cos\theta_\nu\sim 0.3$.

\subsection{Dependence on dimension}
We describe the dynamics of neutrino oscillation by beginning with fiducial model, S1M1. We then discuss the effects of spatial dimension and momentum space dimension.

\subsubsection{fiducial model}
Let us see the fiducial model, S1M1, which is performed in one spatial dimension, $z$,  and one momentum dimension $\theta_\nu$. This setup is called a one-dimensional setup.
The range of the spatial coordinate, $z$, covers $-10\,{\rm cm}\le z \le 10\,{\rm cm}$. In our study, we impose a periodic boundary condition as Refs.~\cite{Bhattacharyya2021,Meng-Ru2021}. As written in Table~\ref{tb:models2_2}, the grid number for $z$ is $N_z=1024$ and the grid number for $\theta_\nu$ is $N_{\theta_\nu}=200$. As in S22~\cite{Sasaki2022}, we use Gauss-Legendre grid to minimize the error in the integration of $\theta_\nu$. The self-interaction Hamiltonian, Eq.~\eqref{eq:hvvM1}, is characterized by $\mu=1.93\times10^{13}\,{\rm s}^{-1}$. The average energy $\langle E_{\nu} \rangle$ in Table~\ref{tab:phys} is calculated from Eq.~\eqref{eq:averagedenergy} and is used for Eq.~\eqref{eq:hvac}.
%The average energy is calculated by Eq.~\eqref{eq:averagedenergy} and is used in Eq.~\eqref{eq:hvac}, see Table~\ref{tab:phys}.

\begin{figure} [htbp]
\includegraphics[width=.95\linewidth]{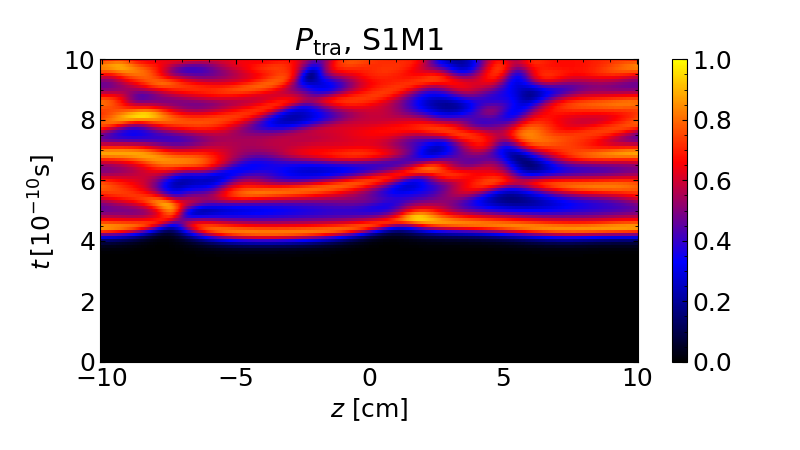}
\caption{Transition probability of S1M1 model is written in space-time diagram.}\label{fig:t-z-PtraS1M1}
\end{figure}

In order to investigate the dynamics of flavor conversions, we calculate the evolution of the transition probability,
\begin{equation}
 P_{\rm tra}= \frac{\langle \rho_{ee,{\rm ini}} \rangle -\langle \rho_{ee} \rangle}{ \langle \rho_{ee,{\rm ini}} \rangle - \langle \rho_{xx,{\rm ini}}\rangle}, \label{eq:Ptra}
\end{equation}
where $\langle A \rangle$ is a average of the quantity $A$ and $\rho_{ee(xx)}$ is an initial value of the diagonal component of the neutrino density matrix. This average is carried out over the neutrino momentum $(E_\nu, \theta_{\nu}, \phi_{\nu})$ and the spatial coordinate $z$ that are variables characterizing the neutrino conversion models in Tables~\ref{tb:models} and~\ref{tb:models2_2} . For S1M1 model, the average can be done over the coordinates $(\theta_{\nu},z)$.
Figure~\ref{fig:t-z-PtraS1M1} shows the spatial and time evolution of transition probability $P_{\rm tra}(z;t)$ in S1M1 model. This quantity is calculated from Eq.~(\ref{eq:Ptra}) with the average over the scattering angle $\theta_{\nu}: \langle A\rangle = \int {\rm d}\cos\theta_\nu A/2$. The transition probability significantly increases at $t=4\times 10^{-10}\,{\rm s}$ irrespective of the value of $z$. The instability grows everywhere at the same time because a random perturbation is imposed on initial values of neutrino density matrices everywhere in $-10\,{\rm cm}\le z \le 10\,{\rm cm}$. From this significant increase of the transition probability, the growth rate for the FFC is estimated as
$O(10^{-10})\,{\rm s}$, which are consistent with the value obtained from our previous linear stability analysis (see Figure~4 of D20~\cite{PhysRevD.101.023018}). After $t=4\times 10^{-10}\,{\rm s}$, the transition probability in Figure~\ref{fig:t-z-PtraS1M1} oscillates for several cycles with the timescale of $t=O(10^{-10})\,{\rm s}$ at all points of $z$. From the value of $\mu$ in Table~\ref{tab:phys}, such oscillation time scale corresponds to $O(100)\mu^{-1}$ which is consistent with the oscillation time scale after the instability saturation as found in previous works (e.g., see Figure~1 of Ref.~\citep{Richers2022}).

\begin{figure} [htbp]
\includegraphics[width=.5\linewidth]{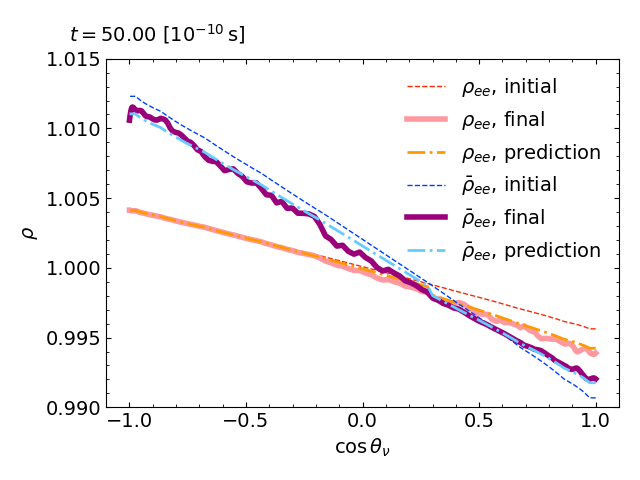}\includegraphics[width=.5\linewidth]{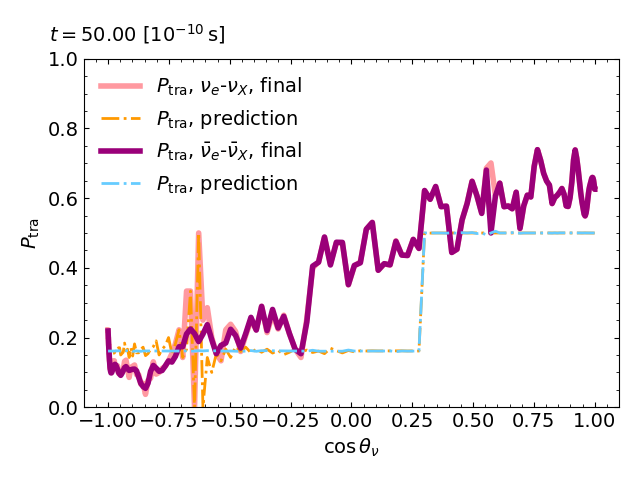}
\caption{Left: The initial and final states of the density matrix of S1M1 model are shown as a function of $\cos\theta_\nu$.
$\rho_{ee}$ (dashed red) and $\bar{\rho}_{ee}$ (dashed blue) correspond to the density matrix of initial $\nu_e$ and $\bar\nu_e$, respectively (same as Figure~\ref{fig:the-rho_ini}).
The final states of $\rho_{ee}$ and $\bar{\rho}_{ee}$ are indicated by pink and purple solid lines, respectively. The theoretical prediction is given by orange and dashed-dotted sky blue line for $\nu_e$ and $\bar\nu_e$, respectively.
Right: The transition probability is shown. The color of the line is the same as that of the left panel.
}\label{fig:the-rho_fin}
\end{figure}

After $t=10\times 10^{-10}\,{\rm s}$, the oscillation state reaches equilibrium.
The left panel of Figure~\ref{fig:the-rho_fin} shows the initial and final profile of the density matrix, the snapshot of $t=50\times 10^{-10}\,{\rm s}$. In the region of $0.3\le\cos\theta_\nu\le1$, $\rho_{ee}$ decreases and $\bar{\rho}_{ee}$ increases and the initial difference of density matrix is mitigated. A series of works is devoted to predict the asymptotic state of the neutrino \cite{Dasgupta2018,Bhattacharyya2020,Bhattacharyya2021,Zaizen2023a,Zaizen2023b,Xiong2023b}. In our cases, we can use the formula for periodic boundary, i.e., Eqs.~(22) and (23) of Ref.~\cite{Zaizen2023a}.
We plot the theoretical prediction in orange ($\nu_e$) and sky blue ($\nu_x$) curves in Figure~\ref{fig:the-rho_fin}.  Our final state roughly agrees with the prediction.  In the region of $0.3\le\cos\theta_\nu\le1$, the survival probability is $\sim 0.5$ and the survival probability of other region is determined by the lepton number conservation. In the right panel, we explicitly show the transition probability using the spatial average, $\langle A \rangle =\frac{1}{z_{\rm max}-z_{\rm min}}\int {\rm d}z A$ in Eq.\eqref{eq:Ptra}.
The fluctuation pattern changes by time. We may obtain a smoother profile using time averaging.
The regions of $0.3\le\cos\theta_\nu $ and $\cos\theta_\nu\le -0.25$ agree with the prediction. Note that  at $\cos\theta_\nu \sim -0.7$, the denominator in Eq.~\eqref{eq:Ptra}, $\langle \rho_{ee} \rangle_{\rm ini} -  \langle \rho_{xx}\rangle_{\rm ini}$, becomes zero (see Figure~\ref{fig:the-rho_ini}) and 
we cannot define the transition probability and numerical round-off error appears in the panel. In the middle region, the theoretical prediction jumps continuously from left to right at $\cos\theta_\nu = 0.3$, but the simulation result is smooth, connecting the two regions. We may need a more detailed formula for a more precise prediction \cite[e.g.,][]{Xiong2023b}.

Note that the time evolution would depend on the choice of boundary conditions and types of perturbations. Ref.~\cite{Meng-Ru2021} compares the random perturbation with point-source-like ones. They found that the averaged final asymptotic state does not strongly depend on the types of perturbation though the spatial pattern and the state before the saturation are different.
In our simulation, the spatial grid covers from $-10$\,{\rm cm} to $10$\,{\rm cm}, the light crossing time is $\sim 3.3 \times 10^{-10}$\,{\rm s}. After that time scale, the results can be affected by the boundary.
However, the interaction of flavor waves should occur before that timescale, since we impose random perturbation on each spatial grid. In that sense, we believe that the simulation results do not strongly depend on the size of the simulation box. In this paper, our aim is to compare the 1D and 0D results. For this purpose, the random perturbation with such a box size is preferable. We do not employ other box size and type of boundary condition.

\begin{figure} [h]
\includegraphics[width=.9\linewidth]{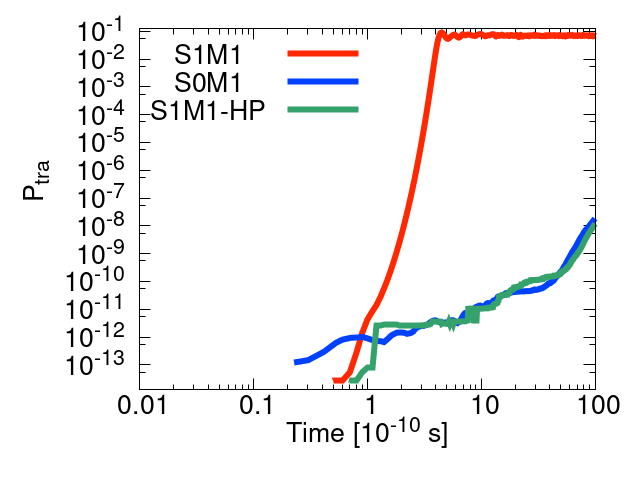}
\caption{Comparison of S0M1, S1M1, and S1M1-HP in the time evolution of transition probability.}\label{fig:t-PtraHovsIHlog}
\end{figure}

\begin{figure} [htbp]
\includegraphics[width=.5\linewidth]{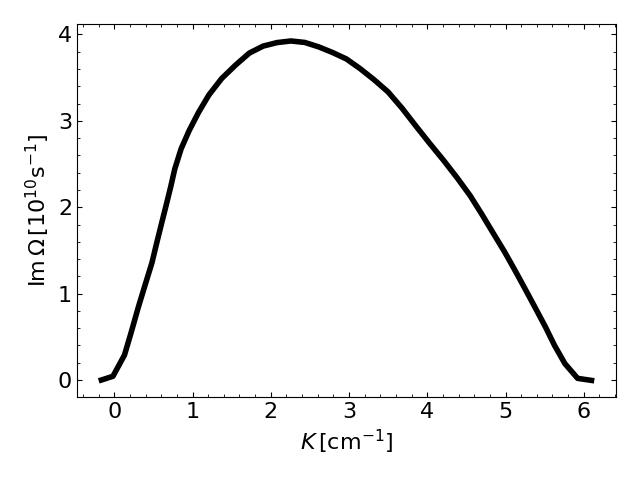}\includegraphics[width=.5\linewidth]{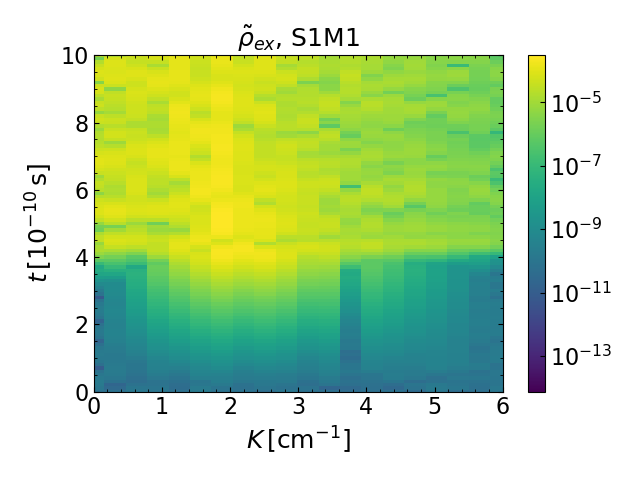}
\caption{
Left: The dispersion relation of FFC, which is similar to the one we obtained in D20~\cite{PhysRevD.101.023018}.
The peak of the growth rate, $\Omega_{\rm peak}$ is $3.9\times 10^{10}\,{\rm s^{-1}}$, see Table~\ref{tab:phys}.
Right: The evolution of the Fourier component of ${\rm Re}(\rho_{ex})$ in the S1M1 model.}
\label{fig:wavenumber}
\end{figure}

\subsubsection{spatial dimension} 

To understand the effect of spatial dimension, we compare the 1D model with the 0D model. As mentioned before, 0D means the advection term in Eqs.~\eqref{eq:QKEneu} and \eqref{eq:QKEant} is ignored.

The transition probability depends on the spatial dimension and the shape of the perturbations.
Figure~\ref{fig:t-PtraHovsIHlog} shows the time evolution of the transition probability of S0M1, S1M1, and S1M1-HP.
In S1M1 and S1M1-HP, we evaluate the spatial average of the probability
where the spatial and angle average are taken in $\langle A \rangle$ in Eq.~\eqref{eq:Ptra}. 

The growth of the instability in the 0D model (blue) is significantly slower, 
$\Omega_{\rm 0D}\sim 8\times 10^{8}\,{\rm s^{-1}}$, than that of 1D model (red), $\Omega_{\rm 1D}\sim 8\times10^{10}\,{\rm s^{-1}}$ in Figure~\ref{fig:t-PtraHovsIHlog}, which is summarized in Table~\ref{tab:phys}.
This is because the growth rate of the instability depends on the wave number and the 0D model does not capture the most rapidly growing mode.
The linear growth rate of this initial condition is evaluated in D20~\cite{PhysRevD.101.023018}.
In the linear analysis, we usually take
\begin{equation}
\rho = \frac{\rho_{ee, {\rm ini}}+\rho_{xx, {\rm ini}}}{2}+\frac{\rho_{ee, {\rm ini}}-\rho_{xx, {\rm ini}}}{2}\left(
\begin{array}{cc}
    s &  S\\
    S^{*} &-s 
\end{array}
\right),
\end{equation}
where $|s|^2+|S|^2=1$.
The evolution of the non-diagonal component of density matrix, $S$, is proportional to $e^{-i(\Omega t-\bf{K}\cdot\bf{x})}$, which is associated with the wave number vector $\bf{K}$. 
Taking $K$ as the wave number of the radially outward direction, we show the dispersion relation in the left panel of Figure~\ref{fig:wavenumber}. 
From the magenta curve in Figure~4 of D20~\cite{PhysRevD.101.023018}, we need to shift the wave number following $-k=K-\Phi$, and $\Phi = \int \frac{{\rm d}\cos\theta_{\nu}}{2}G_{\bm{v}}\cos\theta_{\nu}$ where $k$ is the wave number of the radially inward direction of neutrinos and $G_{\bm{v}}$ is the ELN angular distribution including neutrino distributions (see D20~\cite{PhysRevD.101.023018} for more details). From the initial neutrino distributions, we can calculate $G_{\bm{v}}$ and obtain $\Phi\sim 1.5\,{\rm cm}^{-1}$.

In the 0D setup, only the $K=0$ mode can be captured (it is called homogeneous mode \cite{Dasgupta2018}). In the left panel of Figure~\ref{fig:wavenumber}, the growth rate at $K=0$ is significantly lower than the most rapidly growing mode at $K\sim 2\,{\rm cm}^{-1}$.
This is the reason why the 0D model shows a very slow growth compared to the 1D model.

On the other hand, the 1D setup allows the finite value of $K$ and the mode with $1\,{\rm cm}^{-1}<K<5\,{\rm cm}^{-1}$ can grow relatively rapidly. This is confirmed in the right panel of Figure~\ref{fig:wavenumber}, which shows the Fourier component of ${\rm Re}(\rho_{ex})$ as a function of time and wave number in the S1M1 model.  In the linear growth phase ($t<4\times10^{-10}\,{\rm s}$), 
the mode with $1\,{\rm cm}^{-1}<K<5\,{\rm cm}^{-1}$ grows (yellow). In the non-linear phase ($t>4\times10^{-10}\,{\rm s}$), the Fourier component extends more widely. Such a widely spread spectrum is also found in Ref.~\cite{Richers2022}, see their Figure~3.
On the growth rate, the most rapidly growing mode is $\Omega_{\rm peak}$ is $3.9\times 10^{10}\,{\rm s^{-1}}$ in Figure~\ref{fig:wavenumber} and is consistent with one in the simulation, $\Omega_{\rm 1D} \sim 8\times 10^{10}\,{\rm s^{-1}}$.
Note that transition probability in linear phase is written as $P_{\rm tra}\sim\frac{1}{4}|S|^2 \propto \exp(2\Omega_{\rm peak} t)\sim \exp(\Omega_{\rm 1D} t)$.

The previous interpretation is confirmed in more detailed analysis. S1M1-HP in Figure~\ref{fig:t-PtraHovsIHlog} shows the result of the same 1D model as S1M1 while using the different spatial perturbation at $t=0\,{\rm s}$ i.e., 
\begin{align}
    \rho_{ex}(\theta_\nu,z)=& \rho_{ex}(\theta_\nu,0),\label{eq:S1M1Hpn}\\
    \bar{\rho}_{ex}(\theta_\nu,z) =&  \bar{\rho}_{ex}(\theta_\nu,0),\label{eq:S1M1Hpnb}
\end{align}
which mimics the 0D. The same perturbation is used for all positions. The growth of the transition probability in S1M1-HP (green) is similar to that of S0M1 (blue) after $t=10^{-10}\,{\rm s}$ and significantly slower than that of S1M1 (red) due to the homogeneous perturbation in Eqs.~\eqref{eq:S1M1Hpn} and ~\eqref{eq:S1M1Hpnb}.
In Table~\ref{tb:models2_2}, we summarize the results of the transition probability and write "Suppressed" in the line of S1M1-HP.

\begin{figure} [h]
\includegraphics[width=.9\linewidth]{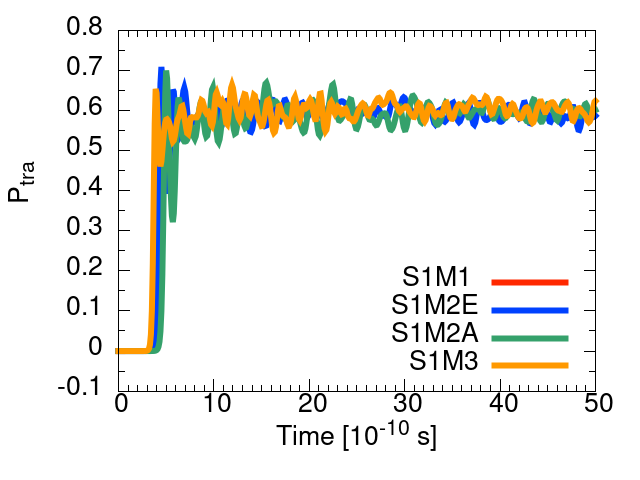}
\caption{Dependence of the momentum space dimension in the time evolution of the transition probability. We compare S1M1\,(red), S1M2E\,(blue), S1M2A\,(green), and S1M3\,(orange) to see the dependence of the momentum space dimension in 1D models.}\label{fig:t-PtraIHDim}
\end{figure}

\subsubsection{momentum space dimension} 

In 1D setup, the dimension of momentum space does not significantly change the transition probability.
Figure~\ref{fig:t-PtraIHDim} shows the time evolution of the transition probability of S1M1, S1M2E, S1M2A, and S1M3.
The definition of the transition probability is similar to Eq.~\eqref{eq:Ptra} but the averaging depends on the dimension. In Table~\ref{tb:models2_2}, we summarize the results of transition probability and note "Not affected" in the line of these models. 

\begin{figure} [htbp]
 \includegraphics[width=.5\linewidth]{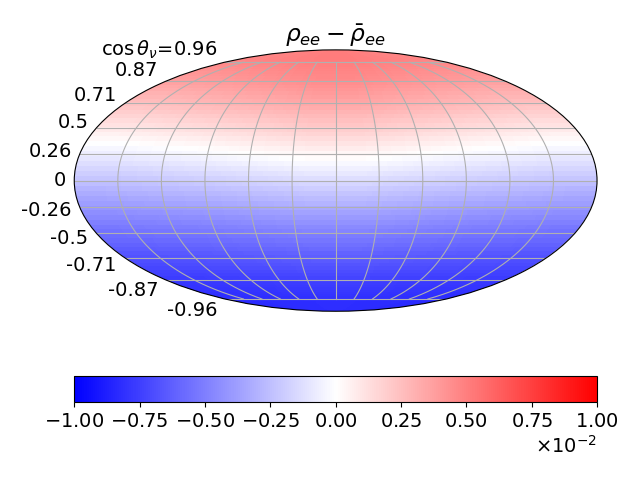}\includegraphics[width=.5\linewidth]{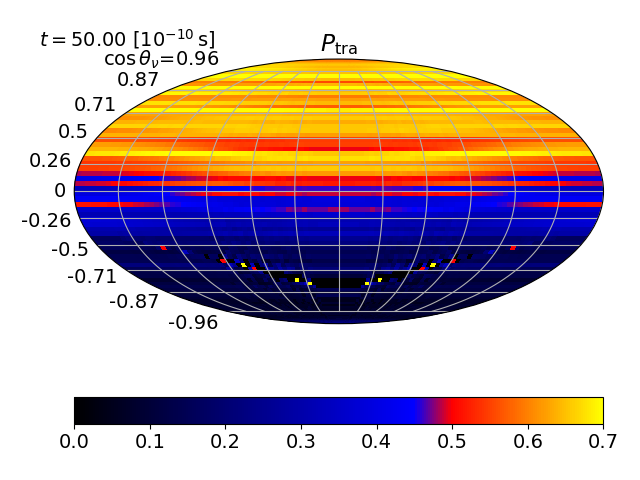}
 \caption{Left: The initial profile of S1M2A. Initial $\rho_{ee}-\bar{\rho}_{ee}$ is displayed in Mollweide projection. Right: The transition probability of the final profile in S1M2A.  
 In both panels, the north pole corresponds to $\theta_\nu=0$, and the vertical ticks are written in $\cos\theta_\nu$. The west and east ends correspond to $\phi_\nu=0$ or $2\pi$. }\label{fig:the-phi-eln}
\end{figure}

One spatial and one momentum (1+1) dimensions may be enough to capture the feature of the instability for our initial profile. The left panel of Figure~\ref{fig:the-phi-eln} shows $\rho_{ee} - \bar{\rho}_{ee}$ of S1M2A at the initial phase. That is almost axisymmetric. The red region ($\rho_{ee} - \bar{\rho}_{ee}>0$) and blue region ($\rho_{ee} - \bar{\rho}_{ee}<0$) is divided by $\cos\theta_\nu \sim 0.3$, which is similar to that of angle averaged profile
(see Figure~\ref{fig:the-rho_ini}).
In the right panel of Figure~\ref{fig:the-phi-eln}, the final transition probability is shown.
The profile is almost axisymmetric and is not so different from the angle averaged profile (see Figure~\ref{fig:the-rho_fin}).
Ref.~\cite{Richers2021} also shows that the spatial dimension does not change the transition probability in their fiducial axisymmetric setup.
If the initial profile deviates from axisymmetry, the results would depend on the spatial and momentum space dimensions.
Looking to the left panel of Figure~\ref{fig:the-phi-eln} more carefully, we found slight azimuthal dependent structures, $\rho_{ee} - \bar{\rho}_{ee}=0$ near $(\cos\theta_\nu, \phi_\nu) = (0,\pi)$, and that does not disappear in the right panel.

\begin{figure} [h]
 \includegraphics[width=.9\linewidth]{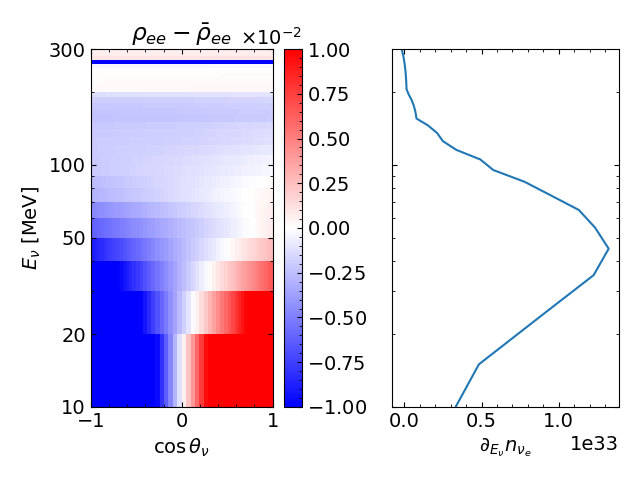}
 \caption{The initial profile of M2E simulations. Left: The initial $\rho_{ee}-\bar{\rho}_{ee}$ is displayed as a function of $E_\nu$ and $\cos\theta_\nu$. Here, the color bar scales as $10^{-2}$.  Right: The number spectrum of electron-type neutrino, $\frac{\partial n_{\nu_e}}{\partial E_\nu}$. The vertical axis is the same as the left panel, $E_\nu$. }\label{fig:the-ene-eln}
\end{figure}

\begin{figure} [h]
 \includegraphics[width=.9\linewidth]{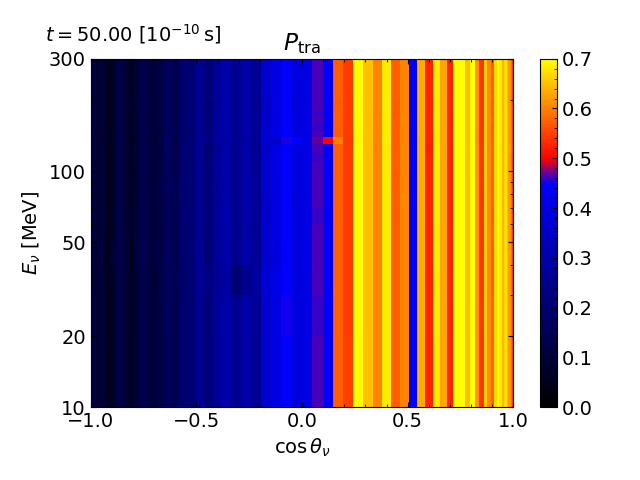}
 \caption{The final transition probability of S1M2E. The probability is written as a function of $\cos\theta_\nu$ and $E_\nu$.}\label{fig:the-ene-Ptr}
\end{figure}

From the result of S1M2E, we conclude that the averaged properties are more critical to determine the final phase than the detailed energy-dependent properties. 
Our initial condition of S1M2E depends on the neutrino energy and the left panel of Figure~\ref{fig:the-ene-eln} shows the initial ELN crossing profiles.
In the right panel, the spectrum $\frac{\partial n_{\nu_e}}{\partial E_\nu}$, is shown and it is defined as 
$\frac{\partial n_{\nu_e}}{\partial E_\nu} = \iint \frac{E^2_\nu{\rm d}\Omega_\nu}{(2\pi\hbar c)^3}f_{\nu_{e}}$.
The peak of spectrum 
is $\sim50\,{\rm MeV}$ and the ELN crossing appears in $\cos\theta_\nu\sim 0.3$, which is consistent with
the energy-integrated profile  in Figure~\ref{fig:the-rho_ini}.
The position of ELN crossing in the low-energy region ($E_\nu<20\,{\rm MeV}$) is almost $\cos\theta_\nu\sim 0$ and there is no ELN crossing in the high-energy region ($E_\nu>100\,{\rm MeV}$). These are different from the averaged profile, where ELN crossing appears in $\cos\theta_\nu\sim 0.3$ (see Figure~\ref{fig:the-rho_ini}). These deviations from the averaged profile do not significantly change the final transition probability. Figure~\ref{fig:the-ene-Ptr} shows the transition probability as a function of $\cos \theta_\nu$ and $E_\nu$. The probability does not depend on the energy and that in $0.3 < \cos\theta_\nu < 1$ becomes equilibrium, $\sim 0.5$. This  feature is similar to that of energy-integrated profile (see Figure~\ref{fig:the-rho_fin}).
\begin{figure} [h]
\includegraphics[width=.9\linewidth]{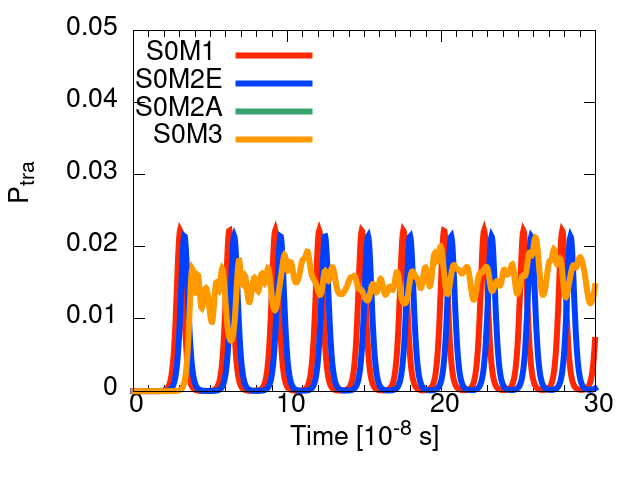}
\caption{Dependence of the momentum space dimension in time evolution of transition probability. We compare S0M1~(red), S0M2E~(blue), S0M2A~(green), and S0M3~(orange) to see the dependence in 0D models.
S0M2A~(green) overlaps S0M3~(orange) and is not visible in the figure.
}\label{fig:t-PtraHDim}
\end{figure}

In 0D setup, the non-axisymmetric motion significantly alters the transition probability.
Figure~\ref{fig:t-PtraHDim} shows the time evolution of the transition probabilities of S0M1, S0M2E, S0M2A, and S0M3. The periodic motion of S0M1 is associated with the axisymmetric feature (only $\theta_{\nu}^\prime$ dependence) in Eq.~\eqref{eq:hvvM1} and is consistent with the previous 0D calculations \cite{Shalgar2021,Sasaki2022,Kato2022}. S0M2E shows periodic motion similar to S0M1 because the fast oscillation mode without collision is induced by the dependence of the neutrino angular distribution and the $E^{\prime}_{\nu}$ dependence in Eq.~\eqref{eq:hvvM2E} has minor effects on the fast mode although such energy dependence is the origin of the slow mode~\cite{Duan2010}. On the other hand, S0M2A breaks the periodicity and significantly alters the flavor conversion.
Such a broken periodicity is induced by non-axisymmetric term $\phi^{\prime}_\nu$ in Eq.~\eqref{eq:hvvM2A} and is consistent with the result in Ref.~\cite{Shalgar2022}. 
The evolution of S0M3 is same as S0M2A. In Table~\ref{tb:models}, we note "Break periodicity" in S0M2A and S0M3. In the absence of collision, the fast flavor conversions in the 0D model are significantly affected by the $\phi_\nu$ dependence although such non-axisymmetric effect is smeared out in the 1D model. Such feature is also confirmed in the flavor conversions involving collision terms as shown in the next section.\\

\subsection{Effect of collision}

\begin{figure} [htbp]
 \includegraphics[width=15cm]{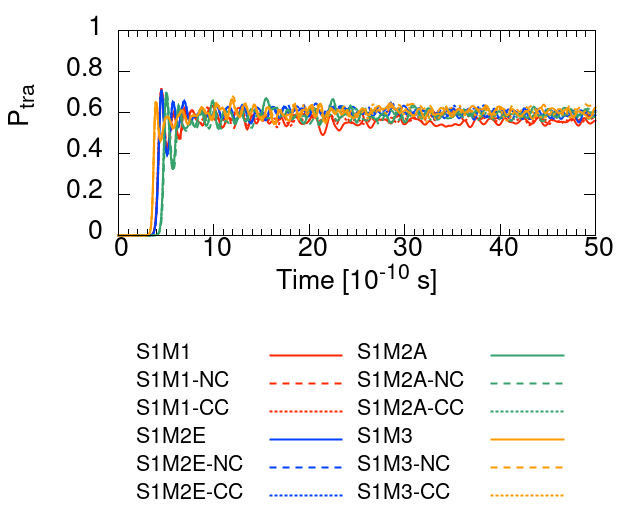}
 \caption{Effect of the scattering in 1D setup. The evolution of transition probability of 12~models are compared. The models with four types of momentum space dimensions are displayed, i.e., S1M1, S1M2E, S1M2A and S1M3. In these models, we add two types of collision, -NC and -CC.}\label{fig:t-PtraS1M2ALLcol}
\end{figure}

Collision does not strongly change the transition probability in 1D setup.
Figure~\ref{fig:t-PtraS1M2ALLcol} shows the time evolution of the transition probability of 12 one-dimensional models. The four types of momentum dimensions (M1, M2A, M2E, M3) and three collisions (none, -NC, -CC) are considered. We do not observe any significant enhancement or suppression compared to S1M1.
These results are consistent with Refs.~\cite{Martin2021,Sigl2022}. They also claim that the collision does not affect the transition probability. In this study, this fact is confirmed more systematically. 
In Table~\ref{tb:models2_2}, we note that "Not affected" in the line of these models. 
It is worth mentioning that Refs.~\cite{Martin2021,Sigl2022} employs a density gradient that is not the same as in our setup.

\begin{figure} [htbp]
 \includegraphics[width=15cm]{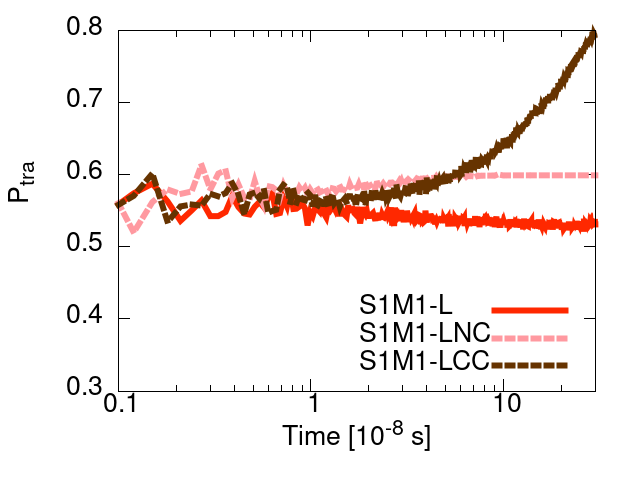}
 \caption{A long-term evolution of transition probability in 1D setup with scattering.  S1M1-L, S1M1-LNC, and S1M1-LCC are colored in red, pink, and brown, respectively.
 }\label{fig:t-PtraS1Lcol}
\end{figure}

An interesting feature is found in a long-term simulation.
Figure~\ref{fig:t-PtraS1Lcol} compares the transition probability of S1M1-L, S1M1-LNC, and S1M1-LCC. The simulation time is extended to $t=30\times 10^{-8}\,{\rm s}$ (the fiducial model, S1M1, is carried out until $t=1\times 10^{8}\,{\rm s}$).
We found a slight increase in transition probability in S1M1-LNC (pink line).
Here $\kappa_0\sim 0.43\times 10^{8}\,{\rm s}^{-1}$ and $\kappa_1\sim 0.17\times 10^{8}\,{\rm s}^{-1}$ obtained in Eq.~\eqref{eq:kappa0}. 
%compare other rates in Table~\ref{tab:phys}.
%Compare it to the other instability in Table~\ref{tab:phys}.
The effect of the collision appears after $t \ge 10^{-8}\,{\rm s}$.
At the final epoch ($t\sim30\times 10^{-8}\,{\rm s}$), the angular distributions of neutrinos in S1M1-LNC become completely flat. It is worth repeating that this initial condition is taken from the radius of 19\,km, which is inside the neutrino sphere. The isotropic angular distribution can be achieved in this environment. 
Note that the neutrino is moving $\sim 0.09\,{\rm km}$ during $30\times 10^{-8}\,{\rm s}$ and the matter density and collision rate become smaller (larger) during outward (inward) propagation.
To make the setup realistic, we may need to employ the Dirichlet boundary condition \cite{Nagakura2022GRQKNT,Nagakura2022dec,Nagakura2023mar,Zaizen2023b} instead of the periodic boundary condition, which we currently use. 

We found a significant improvement in probability in S1M1-LCC.
We speculate on the enhancement as a collisional instability \cite{Johns2023} that is triggered by the difference between the collision rate of $\nu_e$ and $\bar{\nu}_e$.
The instability criterion is shown in Eq.(6) in Ref.~\cite{Johns2023} and is written as
\begin{equation}
    \frac{n_{\bar{\nu}_e}-n_{\bar{\nu}_x}}{n_{\nu_e}-n_{\nu_x}} \ge \frac{\bar{\Gamma}^{\rm CC}}{\Gamma^{\rm CC}}.
\end{equation}

The emission and absorption can contribute to this instability but those are ignored in this study (see Ref.~\cite{Kato2023}).
The neutrino density is written as $n_i = n_\nu +\delta n_i$ where $n_\nu=1.0\times 10^{35}\,{\rm cm^{-3}}$.
At $t=1\times 10^{-8}\,{\rm s}$,
$(\delta n_{\nu_e},~\delta n_{\bar{\nu}_e},~\delta n_{\nu_x},~\delta n_{\bar{\nu}_x})
=(1.84,~2.65,~2.0,~3.1)\times 10^{32}\,{\rm cm^{-3}}$,
Those are initially
$(2.44,~4.31,~1.44,~1.44)\times 10^{32}\,{\rm cm^{-3}}$.
$\Gamma^{\rm CC}=0.058\times 10^{8}\,{\rm s^{-1}}$ and $\bar{\Gamma}^{\rm CC}=0.018\times 10^{8}\,{\rm s^{-1}}$, which is significantly lower than the growth rate of FFC, $\Omega_{\rm 1D}$ (see Table~\ref{tab:phys}).
Substituting these values, we found that our system is unstable.

We evaluate the growth rate of the instability in our model at $t=5\times 10^{-8}\,{\rm s}$ in Figure~\ref{fig:t-PtraS1Lcol}, and the rate is $\Omega_{\rm S1M1-LCC}=1.5\times 10^{6}\,{\rm s^{-1}}$. As expected, the growth rate is slower than FFC (see Table~\ref{tab:phys}). 
Theoretically, we can also evaluate the growth rate, $\Omega_{\rm CI}$, using Eq.~(14) in Ref.~\cite{Johns2023} (see also Refs.~\cite{Lin2023,Liu2023}) and
\begin{equation}
\Omega_{\rm CI} \sim \frac{\Gamma^{\rm CC}-\bar{\Gamma}^{\rm CC}}{2}\frac{n_{\nu_e} - n_{\nu_x} + n_{\bar{\nu}_e} - n_{\bar{\nu}_x} }{\left|n_{\nu_e} - n_{\nu_x} - n_{\bar{\nu}_e} + n_{\bar{\nu}_x}\right|}
-\frac{\Gamma^{\rm CC}+\bar{\Gamma}^{\rm CC}}{2}.
\end{equation}
Substituting the values at $t=1\times 10^{-8}\,{\rm s}$, we obtain $\Omega_{\rm CI} \sim 0.406 \times 10^{6}\,{\rm s^{-1}}$. This value is closer to that is observed in our simulation.

\begin{figure} [htbp]
 \includegraphics[width=.9\linewidth]{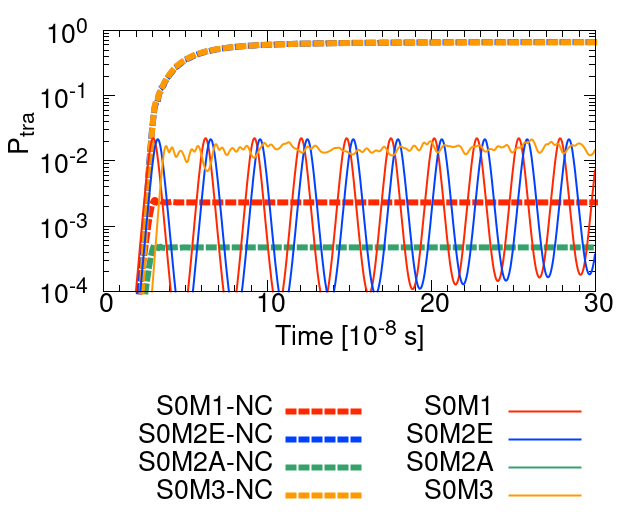}
 \caption{The effect of NC-collision in 0D models. The logarithmic transition probability is shown as a function of time.
 Dashed lines are used for the NC-collision models, whereas thin solid lines are used for the models without collision. Similar to Figure~\ref{fig:t-PtraHDim},  momentum dimension is indicated by color, i.e., S0M1~(red), S0M2E~(blue), S0M2A~(green), S0M3~(orange). S0M2A~(solid green) overlaps S0M3~(solid orange) and is not visible in the figure. Similarly, S0M2E-NC~(dashed blue) overlaps S0M3-NC~(dashed orange) and is not visible in the figure.}

 \label{fig:t-PtraS0NCcol}
 \end{figure}
 \begin{figure} [htbp]
 \includegraphics[width=.9\linewidth]{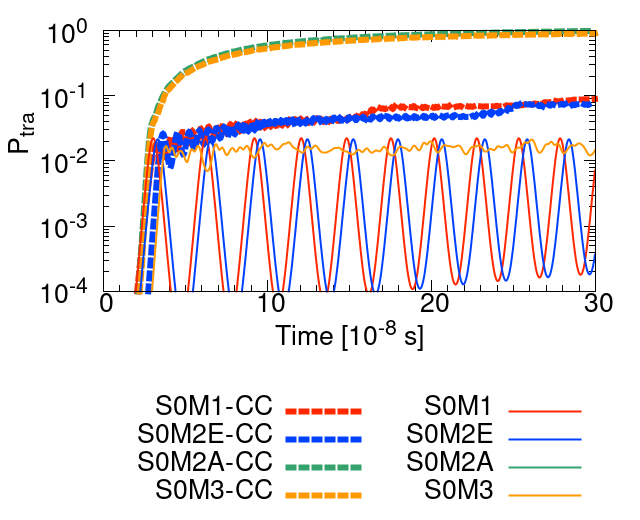}
 \caption{The effect of CC-collision in 0D models.  Same to Figure~\ref{fig:t-PtraS0NCcol}, but the impact of CC-collision is displayed. S0M2A~(solid green) overlaps S0M3~(solid orange) and is not visible in the figure. Similarly, S0M2A-CC~(dashed green) overlaps S0M3-CC~(dashed orange) and is not visible in the figure.}
 
\label{fig:t-PtraS0CCcol}
\end{figure}

In contrast to FFC in 1D setup (see Figure \ref{fig:t-PtraS1M2ALLcol}), the collision strongly changes the transition probability in 0D configuration.
Figures~\ref{fig:t-PtraS0NCcol} and \ref{fig:t-PtraS0CCcol} show the time evolution of the transition probability of models with NC-collision and CC-collision, respectively.
Models without collision are also shown by thin solid lines. The same colors as in Figure~\ref{fig:t-PtraHDim} are used for momentum dimension, i.e., S0M1~(red), S0M2E~(blue), S0M2A~(green), S0M3~(orange).

It is difficult to quantify the effects of NC-collision. The transition probability is shown in Figures~\ref{fig:t-PtraS0NCcol}.
While S0M1-NC and S0M2A-NC show the suppression of the transition probability compared to those without the collision term,
S0M2E-NC and S0M3-NC show the enhancement of the probability.
The flavor conversion of S0M1-NC is more suppressed than that of S0M1. Such suppression from NC-collisions was also confirmed in Ref.~\cite{Sasaki2022} when $\kappa_{0}^{-1}$ and $\kappa_{1}^{-1}$ are comparable to the timescale of FFC, see Table~\ref{tab:phys}. NC-collision does not induce collisional instability. The suppression and enhancement is due to isotropization of angular distribution by the collision \cite{Johns2022}. Similar to our energy-dependent simulations, S0M2E-NC and S0M3-NC,
Refs.~\cite{Kato2022,Chia2023} also report the enhancement of conversion probability.
Ref.~\cite{Kato2022} reports that we need to change the average energy in the collision term to keep the consistency between the energy-dependent and the energy-averaged simulations.
In this study, we calculate $\langle E_\nu\rangle$  following Eq.~\eqref{eq:averagedenergy} and use it in the collision term.

CC-collision basically enhances the transition probability. We summarize our results in Figures~\ref{fig:t-PtraS0CCcol}. 
While S0M1-CC and S0M2E-CC show a slight enhancement, S0M2A-CC and S0M3-CC show a significant increase.
This means that the azimuthal-angle dependence is important in CC-collisions.
As discussed in the long-term simulations of 1D setup (see Figure~\ref{fig:t-PtraS1Lcol}), collisional instability also occurs with FFC. The timescale of the collisional instability is slower than that of FFC but it increases the growth rate in the late epoch.

\section{Summary and Discussion}\label{sec:summary}

We calculate FFCs with collision effects of neutrino scatterings based on the realistic initial profile that is taken from two-dimensional (2D) fully self-consistent Boltzmann-neutrino-radiation-hydrodynamics simulations for the progenitor model of non-rotating 11.2 $\textrm M_{\odot}$ which were performed on the Japanese K-supercomputers at the post-bounce time of $\textrm t_{\textrm {pb}} = 190$ ms.

The effect of spatial dimension is examined by comparing the 1D model with the 0D model. The transition probability depends on the spatial dimension and the shape of the perturbations. The growth of instability in the 0D model is significantly slower than that of the 1D model because the growth rate on the wave number $K$, and the 0D model does not capture the most rapidly growing mode.

The dimension of momentum space does not significantly affect the transition probability in 1D model. We demonstrate that the averaged properties are more critical for determining the final phase than the detailed energy-dependent properties. In our initial condition, all flavors have similar number density and spectrum. Even if a significant transition probability is observed, this hardly affects the explodability and observation. A non-axisymmetric motion significantly alters the transition probability in 0D model. In the absence of collision, the fast flavor conversions in 0D model are strongly affected by the azimuthal angle dependence. However, such a non-axisymmetric effect is smeared out in 1D model.

The collision-induced enhancement occurs on a long time scale in 1D model. Such enhancement does not appear on the short-time scale in 1D model despite its prominence in the 0D model.  We hypothesize that the enhancement is caused by collisional instability caused by the difference between $\nu_e$ and $\bar\nu_e$ collision rates \cite{Johns2023}.

%\textcolor{blue}{This work is based on a single snapshot and further rigorous efforts beyond this work are required for more general and robust conclusions.}
Our current analysis depends on the mean-field approximation, however, the many-body effect should be considered \cite[see][and references therein]{Birol2018,Volpe2023review,Balantekin2023}. For the 1D model, the periodic boundary condition is employed, but the Dirichlet boundary can be considered \cite{Dasgupta2018,Bhattacharyya2020,Bhattacharyya2021,Zaizen2023a,Zaizen2023b,Xiong2023b}. The results depend on the boundary condition, and such a simplified setup could be different from the realistic setup in supernovae~\cite{Johns2022}. We may need further discussion on this issue. So far, we consider neutrino scattering. Emission and absorption~\cite{Kato2023}, investigating multiple snapshots and other processes~\cite{Xiong2023b} should be considered in the future. Even though we used a two-flavor approximation in this paper, extending the formulation to three flavors is certainly necessary~\cite{Chakraborty2020,Shalgar2021jul}. We remark that our FFC calculation is decoupled from the 2D CCSN simulation and the feedback from FFC to hydrodynamics~\cite{Suwa_2011,Ehring2023a,Ehring2023b} is not taken into account. Further development will be needed for self-consistent implementation \cite[e.g.,][]{Nagakura2023may,Xiong2023b}.

%\begin{acknowledgments}
\section*{Acknowledgment}

We thank Shoichi Yamada, Hiroki Nagakura, Chinami Kato, Masamichi Zaizen, Sherwood Richers,
and Lucas Johns for fruitful discussions and useful comments. This study was supported in part by JSPS/MEXT KAKENHI Grant Numbers 
JP21H01083, 
JP23K20848, % Sotani Kiban B, Takiwaki Co-I 
JP23K22494, % Kotake Kiban B
JP23K25895, % Yokoi Kiban B, Takiwaki Co-I 
JP23K03400,  % Takiwaki Kiban C 
and JP24K00631. % Yamamoto Kiban B
%This work was also supported in part by U.S. National Science Foundation Grants PHY-2020275 and PHY-2108339. 
Numerical computations were carried out on GPU and PC cluster at the Center for Computational Astrophysics,
National Astronomical Observatory of Japan.
%This research was also supported by MEXT as “Program for Promoting researches on the Supercomputer Fugaku” (Toward a unified view of the universe: from large scale structures to planets) and JICFuS.
This research was also supported by MEXT as “Program for Promoting researches on the Supercomputer Fugaku” (Structure and Evolution of the Universe Unraveled by Fusion of Simulation and AI; Grant Number JPMXP1020230406) and JICFuS. This work was carried out under the auspices of the National Nuclear Security Administration of the U.S. Department of Energy at Los Alamos National Laboratory under Contract No. 89233218CNA000001. 
%\end{acknowledgments}

\appendix

\section{Normalization of the density matrix}\label{sec:normalization}

As shown in Tables~\ref{tb:models} and ~\ref{tb:models2_2}, we perform a variety of models with different dimensions. The definition of the density matrix and Hamiltonian depends on the dimension of the momentum space. Here we explain how we calculate them.

In general, the density matrix is defined as
\begin{equation}
\rho_{ii}=\left(\int {\rm d}Xf_{i}\right)\left(\int {\rm d}Y\right)/
\left(\int {\rm d}X {\rm d}Y f_{\nu_e}\right),
\end{equation}
where $X,Y$ are some phase space.
%%%%%%%%%%%%%%%%%%%%%%%%%%%%%%%%%%%%%%%%%%%

\subsection{M1 Model} 
In the setup of energy-integrated axisymmetric case (M1), we have
\begin{equation}
{\rm d}X=\frac{E_\nu^2{\rm d}E_\nu{\rm d}\phi_\nu}{(2\pi\hbar c)^3},~
{\rm d}Y={\rm d}\cos\theta_\nu.
\end{equation}
Then we obtain
\begin{align}
\rho_{ii}(\theta_{\nu}) =& 
2\left(\iint \frac{E^2_\nu{\rm d}E_\nu{\rm d}\phi_\nu}{(2\pi\hbar c)^3} f_{\nu_i} \right) /  n_{\nu_e},\\
n_{\nu_e} =& \iint \frac{E^2_\nu{\rm d}E_\nu{\rm d}\Omega_\nu}{(2\pi\hbar c)^3}f_{\nu_{e}}. \label{eq:nnue-m1}%    
\end{align}
Note that our notation assumes $\int \frac{{\rm d}\cos\theta_\nu}{2} \rho_{ee} =1 $ and that would be factor 2 larger than some works that assume $\int {\rm d}\cos\theta_\nu \rho_{ee} =1 $.
The Hamiltonian is written as 
\begin{align}
H_{\nu\nu}(\cos\theta)=\mu
\int^{1}_{-1}\frac{\mathrm{d}\cos\theta^{\prime}_\nu}{2} h_{\nu\nu},\\
h_{\nu\nu}= (\rho(\theta^\prime_\nu)-\bar{\rho}(\theta^\prime_\nu))(1-\cos\theta_\nu\cos\theta^\prime_\nu),\label{eq:hvvM1}
\end{align}
where $\mu=\sqrt{2}G_{F}{n_{\nu_e}}$.
In Eq.\eqref{eq:hvac} and collision terms, we need average neutrino energy. We define it as 
\begin{align}
\langle E_\nu \rangle &= 
\left(\sum_{i}\iiint \frac{E^3_\nu{\rm d}E_\nu{\rm d}\cos\theta_\nu{\rm d}\phi_\nu}{(2\pi\hbar c)^3} f_{\nu_i} \right)\nonumber\\
&/\left(\sum_{i}\iiint \frac{E^2_\nu{\rm d}E_\nu{\rm d}\cos\theta_\nu{\rm d}\phi_\nu}{(2\pi\hbar c)^3} f_{\nu_i} \right),\label{eq:averagedenergy}
\end{align}
where $i=\nu_e$, $\bar{\nu}_e$, $\nu_\mu$, $\bar{\nu}_\mu$.

%%%%%%%%%%%%%%%%%%%%%%%%%%%%%%%%%%%%%%%%%%%
\subsection{M2E Model} 
We call our energy-dependent axisymmetric model as M2E model. The phase space is as follows.
\begin{equation}
{\rm d}X={\rm d}\phi_\nu, {\rm d}Y={\rm d}\cos\theta_\nu.  
\end{equation}
The density matrix for this model is expressed as 
\begin{align}
\rho_{ii}(E_{\nu},\theta_{\nu}) =& 2\left(\int f_{\nu_i}(E_{\nu},\theta_{\nu}, \phi_{\nu}) {\rm d}\phi_{\nu}\right) /  n_{E,\nu_{e}},\\
n_{E,\nu_{e}} (E_{\nu}) =& \left(\int{\rm d}\Omega_\nu f_{\nu_{e}}\right). %    
\end{align}
Then the Hamiltonian is 
\begin{align}
H_{\nu\nu}(\theta_\nu)=&
\int {\rm d}{\mu_E^\prime}
\int^{1}_{-1}\frac{\mathrm{d}\cos\theta^{\prime}_\nu}{2}h_{\nu\nu}.\\
{\rm d}\mu_E= & \sqrt{2}G_F \frac{E^2_\nu{\rm d}E_\nu}{\left(2\pi\hbar c\right)^3} n_{E,\nu_{e}},\\
h_{\nu\nu}=& (\rho(E^\prime_\nu,\theta^\prime_\nu)-\bar{\rho}(E^\prime_\nu,\theta^\prime_\nu))(1-\cos\theta_\nu\cos\theta^\prime_\nu).\label{eq:hvvM2E}
\end{align}

%%%%%%%%%%%%%%%%%%%%%%%%%%%%%%%%%%%%%%%%%%%
\subsection{M2A Model} 
We call our energy-integrated non-axisymmetric model as M2A model. The phase space is as follows.
\begin{equation}
{\rm d}X=\frac{E^2_\nu{\rm d}E_\nu}{\left(2\pi\hbar c\right)^3},~
 {\rm d}Y={\rm d}\cos\theta_\nu{\rm d}\phi_\nu.
\end{equation}
The density matrix for this model is expressed as 
\begin{align}
\rho_{ii}(\theta_{\nu},\phi_{\nu}) =& 
4\pi\left( \int \frac{E^2_\nu{\rm d}E_\nu}{(2\pi\hbar c)^3} f_{\nu_i} \right)
/  n_{\nu_e},\\
n_{\nu_e}=& \left(\int  \frac{E^2_\nu{\rm d}E_\nu {\rm d}\Omega_\nu }{(2\pi\hbar c)^3}
f_{\nu_{e}}\right). %    
\end{align}
 Hamiltonian for this model is expressed as 
\begin{align}
H_{\nu\nu}(\theta_\nu,\phi_\nu)=&
\mu \int \frac{{\rm d}\phi^\prime}{2\pi}\int^{1}_{-1}
\frac{\mathrm{d}\cos\theta^{\prime}}{2} h_{\nu\nu},\\
\mu=&\sqrt{2}G_F n_{\nu_e},\\
h_{\nu\nu} =&  \left[
          \rho(\theta^\prime_\nu,\phi^\prime_\nu)
    -\bar{\rho}(\theta^\prime_\nu,\phi^\prime_\nu)\right]\nonumber\\
&\times\left[1-\cos\theta_\nu\cos\theta^\prime_\nu -\sin\theta_\nu\sin\theta^\prime_\nu\right.\nonumber\\
&\times\left. (\cos\phi_\nu\cos\phi^\prime_\nu+\sin\phi_\nu\sin\phi_\nu^\prime)\right].\label{eq:hvvM2A}
\end{align}
See also Ref.~\cite{Shalgar2022}.
As shown in Figure~\ref{fig:the-phi-eln}, our initial profile slightly depends on $\phi_\nu$ but almost axisymmetric. Averaged energy is as same as Eq.~\eqref{eq:averagedenergy}.

%%%%%%%%%%%%%%%%%%%%%%%%%%%%%%%%%%%%%%%%%%%
\subsection{M3 Model} 
In this model, we can set ${\rm d}X=1$ and 
${\rm d}Y={\rm d}\cos\theta_\nu{\rm d}\phi_\nu$.
Though we write same equations in Section~\ref{sec:formulation}, we repeat it for completeness of this section.

We express our density matrix as
\begin{align}
\rho_{ii}(E_{\nu},\theta_{\nu}, \phi_{\nu}) =& 4\pi  f_{\nu_i}(E_{\nu},\theta_{\nu}, \phi_{\nu})  /  n_{E,\nu_{e}},\\
n_{E,\nu_{e}} (E_{\nu})=& \int {\rm d}\Omega_\nu f_{\nu_{e}}. % 
\end{align}
The Hamiltonian is expressed as 
\begin{align}
H_{\nu\nu}(\theta_{\nu},\phi_\nu)=&
\int {\rm d}{\mu_E^\prime}
\int \frac{{\rm d}\phi_\nu^\prime}{2\pi}\int^{1}_{-1}\frac{\mathrm{d}\cos\theta_\nu^{\prime}}{2}h_{\nu\nu},\\
{\rm d}\mu^\prime_E=&\sqrt{2}G_F \frac{E^{\prime^2}_\nu{\rm d}E_\nu^\prime}{\left(2\pi\hbar c\right)^3}n_{E,\nu_{e}} (E_{\nu}),\\
\begin{split}
h_{\nu\nu} =&[\rho(E^\prime_\nu,\theta_{\nu}^\prime,\phi_{\nu}^\prime)-\bar{\rho}(E^\prime_\nu,\theta_{\nu}^\prime,\phi_{\nu}^\prime)] \\
&\times \left[1-\cos\theta_{\nu}\cos\theta_{\nu}^\prime -\sin\theta_{\nu}\sin\theta_{\nu}^\prime\right.\\
&
\times\left.
(\cos\phi_{\nu}\cos\phi_{\nu}^\prime+\sin\phi_{\nu}\sin\phi_{\nu}^\prime)\right].
\end{split}
\end{align}

\section{Code verification}\label{sec:codeverification}
In this paper, we have updated QDSCNO~\cite{Sasaki2022} which enables us to achieve spatial advection. For the code verification, we have tested the problem in Ref.~\cite{Richers2022} where the code comparison is performed. In the paper, 5 codes are compared, i.e., 
{\small EMU}~\cite{Richers2021emu}, {\small NuGAS}~\cite{Duan2021}, {\small COSE$\nu$}~\cite{George2023cosen}, Bhattacharyya et al~\cite{Bhattacharyya2020}, and Zaizen~\cite{Zaizen2021}.

The initial distribution is 
\begin{align}
    \rho_{ee} &= A \exp(-(\cos\theta_\nu-1)^2/2\sigma_{ee}^2), \label{eq:ini_ee_Richers2022}\\
    \bar{\rho}_{ee} &= \bar{A} \exp(-(\cos\theta_\nu-1)^2/2\bar{\sigma}_{ee}^2), \label{eq:ini_bb_Richers2022}
\end{align}
where $\sigma_{ee}$ and $\bar{\sigma}_{ee}$
 are 0.6 and 0.53, respectively. The normalization constants $A$ and $\bar{A}$ are determined to meet the condition, $\int{\rm d}\cos\theta_\nu \rho_{ee}=1$ and $\int{\rm d}\cos\theta_\nu \bar{\rho}_{ee}=0.9$. The notation is factor 2 different from ours, see the text below Eq.~\eqref{eq:nnue-m1}.
$\rho_{xx}$ and $\bar{\rho}_{xx}$ are 0.
The Hamiltonian is written as 
$H_{\nu\nu}(\cos\theta)=\mu
\int^{1}_{-1}\mathrm{d}\cos\theta^{\prime}_\nu (\rho-\bar{\rho})(1-\cos\theta_\nu\cos\theta^\prime_\nu)$.

\begin{figure} [htbp]
\includegraphics[width=.9\linewidth]{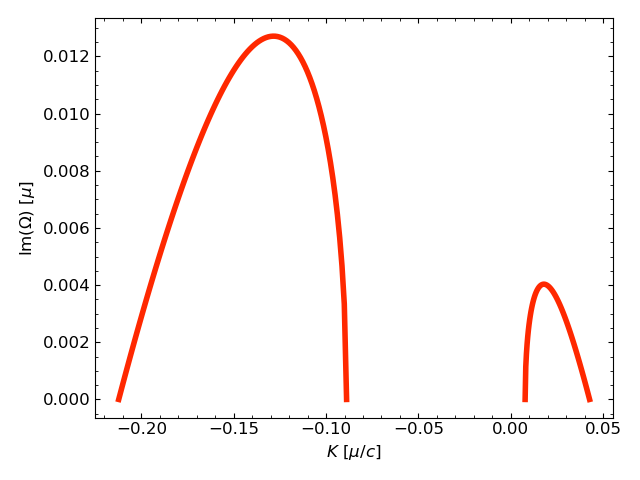}
\caption{The dispersion relation of the initial profile of Eqs.~\eqref{eq:ini_ee_Richers2022} and \eqref{eq:ini_bb_Richers2022}.
}\label{fig:DRRichers2022}
\end{figure}

We calculate the dispersion relation using the code in  {\small NuGAS}~\cite{Duan2021}. The dependence of the wave number on the growth rate is shown in Figure~\ref{fig:DRRichers2022}.
There are two branches of the solution. One appears in $-0.2 \mu/c< K < -0.1 \mu/c$ and the other appears in $0.01  \mu/c< K < 0.04 \mu/c$.
The former one has larger growth rate and its peak is $0.013 \mu$.
The latter one have smaller peak with  $0.004\mu$.
At $K=0$, we do not have unstable mode and we do not expect FFC in the 0D model. 

\begin{figure} [htbp]
\includegraphics[width=.9\linewidth]{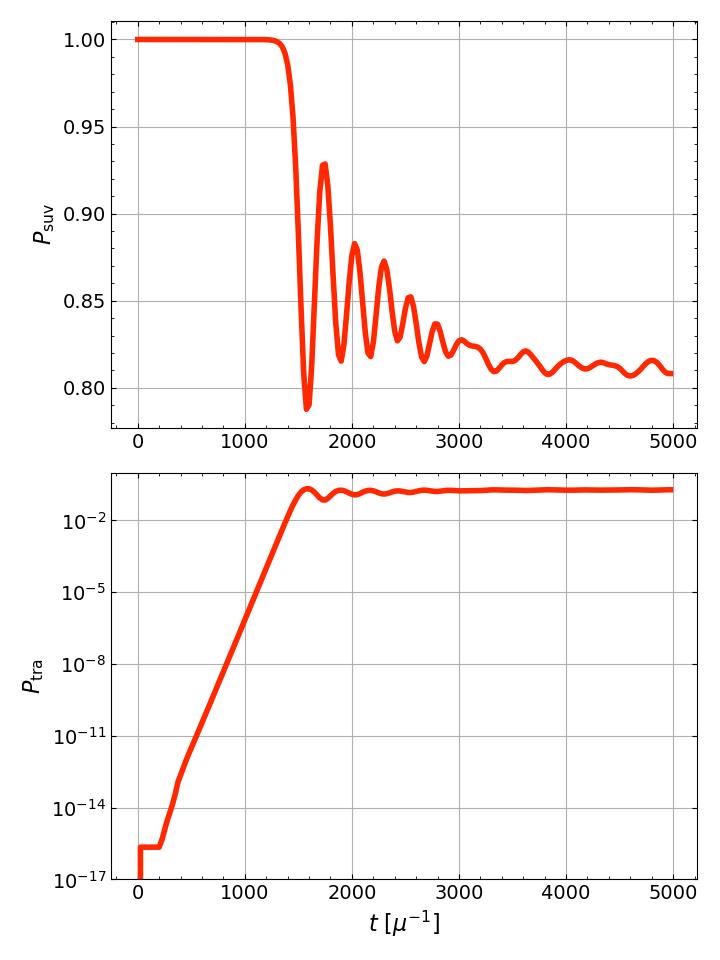}
\caption{The survival probability (top) and the transition probability (bottom) of the test problem in Ref.~\cite{Richers2022}.
}\label{fig:Richers2022}
\end{figure}

In this test, we use $N_{\theta_\nu} = 200$ bins in $\theta_\nu$ with Gauss-Legendre grid.
We adopt a simulation box of size \mbox{$L = 10240 c/\mu$} spanned by a uniform grid of $N_z = 10240$ cells.
We take the unit of $c=1$ and $\mu=1$.
The initial perturbation is the same as Eqs.~(7) and (8) in Ref.~\cite{Richers2022}.

In Figure~\ref{fig:Richers2022}, we show the survival probability (top) and the transition probability (bottom).
The curves are very close to Figure~1 in Ref.~\cite{Richers2022}. In the non-linear phase, the first peak comes at $t\sim1600/\mu$ and the final survival probability is $\sim 0.81$ (see the top panel). The evolution in linear growth phase is also very similar to Figure~1 in Ref.~\cite{Richers2022} (see the bottom panel). From the dispersion relation, the most rapidly growing mode is Im($\Omega$)$\sim 0.013\mu$ at $K=-0.13\mu/c$. The linear growth rate estimated from the evolution of $P_{\rm tra}$ is $\Omega \sim 0.026\mu$.
The growth rate in $P_{\rm tra}$ is generally factor 2 higher than that in the dispersion relation.
The simulation results are consistent with the dispersion relation.
See also Figure~\ref{fig:wavenumber} and the main text for the details.

\bibliographystyle{ptephy}
\bibliography{Papers_vff, Papers_arXiv}

\end{document}